\documentstyle[11pt,epsf]{article}
\def\lsim{\mathrel{\rlap{\lower4pt\hbox{\hskip1pt$\sim$}}
    \raise1pt\hbox{$<$}}}         
\def\gsim{\mathrel {\rlap{\lower4pt\hbox{\hskip1pt$\sim$}}
    \raise1pt\hbox{$>$}}}         

\begin{document}

\par
\topmargin=-1cm      
\vspace*{.5in}

{ \small
\noindent{U.Md. PP\# 99-117}\hfill{DOE/ER/40762-174\\
\noindent{U.W.  PP\# NT@UW-99-19}}}

\vspace{40.0pt}

\begin{center}
  {\large {\bf Deuteron electromagnetic properties
      and the viability of effective field theory methods in the
      two-nucleon system}}

\vspace{18pt}
{\bf D.~R.~Phillips} \footnote{Email: phillips@phys.washington.edu}

\vspace{6pt}
{Department of Physics, University of Washington,
Box 351560, Seattle, WA, 98195-1560,}

\vspace{6pt}
and 

\vspace{6pt}
{Department of Physics, University of Maryland, College Park, MD, 
20742-4111}

\vspace{12pt}
{\bf T.~D.~Cohen} \footnote{Email: cohen@physics.umd.edu}

\vspace{6pt}
{Department of Physics, University of Maryland, College Park, MD, 
20742-4111}
\end{center}

\vspace{12pt}

\begin{abstract} 
  The central tenet of effective theory is that the details of
  short-distance physics will not have a significant impact on
  low-energy observables.  Here we perform an analysis of
  electron-deuteron scattering at low momentum transfers which is
  based on effective field theory.  We show that in our approach the
  deuteron electromagnetic form factors $F_C$ and $F_M$ indeed are
  largely insensitive to the short-range $NN$ potential at momenta $Q$
  up to about 700 MeV. We also find that the effective field theory
  approach to deuteron electromagnetic structure provides a systematic
  justification for many features which have been seen in potential
  model calculations of the same quantities.
\end{abstract}

\newpage

\section{Introduction}

The physics of light nuclei would seem to be fertile soil for the application
of effective field theory techniques. These techniques rely on the existence of
a hierarchy of scales. Once such a hierarchy is established the central tenet
of effective field theory (EFT) is that the physics at energies much lower than
some scale $\Lambda$ cannot depend on the details of the dynamics at that
scale.  The physics at these higher energies appears implicitly in the
low-energy effective theory, encoded in a series of contact operators.  Suppose
then that there are a succession of scales in the problem: 

\begin{equation}
  \Lambda_1 < \Lambda_2 < \Lambda_3 \ldots.
\end{equation} 
If we gradually increase the energy being studied we can imagine
modifying the effective theory to explicitly include the
new physics which emerges as these new scales $\Lambda_n$ are
successively reached.

Light nuclei appear to be an example of a system in which this
kind of scale-separation exists.  First, it is important to note that
in light nuclei the scale of binding is ``unnaturally'' small.  For
instance, poles in the two-nucleon T-matrix occur at momenta
of 8 MeV (in the ${}^1S_0$ channel) and 46 MeV (in the
${}^3S_1-{}^3D_1$ channel). Both of these are well below the
next-lightest scale in the problem, $m_\pi$. Thus, key scales in these
systems include $\sqrt{MB}$, the characteristic momentum scale for
nuclear binding; $m_\pi$, the pion mass; and $\Lambda_\chi$, the scale
of chiral symmetry breaking.  In the $A=2$ system we certainly have
$\sqrt{MB} \ll m_\pi \ll \Lambda_\chi$, although the characteristic
momentum scale inside larger nuclei may be closer to $m_\pi$.

If one confines one's interest to momenta $p \ll \sqrt{MB}$, then an
effective field theory Lagrangian containing contact nucleon-nucleon
interactions with increasing numbers of derivatives can be employed.
This theory has simple power-counting rules since higher-dimensional
operators and loops are suppressed by powers of
$p/\sqrt{MB}$~\cite{Be97B,Ka98B,vK98}.  However, an EFT based on such
counting has a very limited domain of validity. Indeed, it is
tautological that the bound state at energy $B$ cannot itself be
discussed in the context of such an effective theory.  To be useful in
nuclear physics it is essential that the effective theory's domain of
validity be extended well beyond momenta $p$ of order $\sqrt{MB}$ so
that it can describe light nuclei. Somehow the non-perturbative
physics associated with the low-energy poles in the $NN$ t-matrix must
be incorporated in the EFT.

Once that has been achieved we can work in the regime $\sqrt{MB} \sim
p < m_\pi$, and discuss the bound state using a theory which need not
explicitly include even the lightest meson, the pion.  There has been
much recent interest in such pionless effective theories for nuclear
physics, as they serve to elucidate many issues associated with the
treatment of these low-energy bound states in effective field
theory~\cite{Be97B}---\cite{Ch99}.
They certainly simplify calculations, since the only operators which
appear in the effective theory produce separable interactions in the
two-nucleon system.

The pionless theory is of somewhat limited use, since in many
applications momenta $p \sim m_\pi$ are probed. This
``higher-momentum'' regime is, of course, still a low-momentum regime
from the perspective of hadronic physics.  The low-momentum effective
theory of hadronic interactions is chiral perturbation theory, in
which the pions are introduced as explicit fields.  They appear as the
pseudo-Goldstone bosons of the spontaneously-broken (approximate)
chiral symmetry $SU(2)_L \times SU(2)_R$.  A consequence of the
spontaneous breaking of this symmetry is that all pion-pion or
pion-nucleon interactions involve either derivatives or powers of the
pion mass. It follows that a systematic power counting in momenta and
$m_\pi$ may be developed in which higher-order interactions of the
pions are suppressed by powers of $m_\pi$ and/or the typical momentum
scale in the problem, $p$. The techniques of chiral perturbation
theory thus provide a systematic expansion which yields predictions
for the interactions of pions, nucleons and photons at low momenta.
These predictions have been successful in the zero- and one-nucleon
sectors, and are essentially model-independent, as they are based only
on the symmetries of QCD and power counting.  For a detailed 
review on this subject see Ref.~\cite{Br95}.

We would like to also apply these techniques to many-nucleon systems.
However, as noted above, in the two-nucleon case the situation is
complicated by the presence of an additional scale which does not
appear in the one-nucleon case: the characteristic momentum scale of
the bound state, $\sqrt{MB}$. In the $A=2$ case this is very different
from $m_\pi$ and thus is usually treated as a separate scale in EFT
analyses of this problem.  Weinberg's proposal for dealing with the
emergence of this new scale was to apply the methods of chiral
perturbation theory only to graphs in which the energy flowing through
all lines was of order $m_\pi \gg B$~\cite{We90,We91,We92}. For these
graphs the additional low-energy scale $B$ does not result in any
complications, since, by assumption, the energy of all legs is
significantly above that scale.  Thus chiral perturbation theory can
be applied to these "two-particle irreducible" graphs without any
modifications. The two-particle irreducible objects calculated in
chiral perturbation theory must be sewn together with free Green's
functions of the multi-nucleon system in order to generate the actual
S-matrix elements of the theory. (For reviews of the application of
these ideas to the $NN$ potential and probes of light nuclei see
Refs.~\cite{Fr96,vK99}.) It appears then that for the S-matrix
elements themselves no systematic expansion exists, despite the
demonstrable power-counting for objects that appear in their
construction. 

Such a point of view is similar to that seen when
low-energy effective actions are derived using Wilsonian
renormalization group arguments (see Refs.~\cite{Ep98C,Bi98B} for more
explicit investigations of this connection). These techniques are
widely applied, and have, for instance, been used to obtain improved
actions for lattice gauge theory.  In all cases the effective action
is obtained using a perturbation theory in the ratio of low-energy
scales to the Wilsonian RG cutoff scale. That Lagrangian is then
solved to all orders. In the same way chiral perturbation theory can
be used to generate a nonrelativistic, particle-number-conserving
Hamiltonian for the nuclear system.  This Hamiltonian is valid up to a
given order in the chiral expansion and can be employed in any
many-nucleon system at low energy. Once written down it should be
solved to all orders. The interaction of the nucleus with
electromagnetic, weak, and pionic currents can then also be derived in
chiral perturbation theory (see, for instance,
Refs.~\cite{Be97,Be99,Pa98B}). For full consistency the expansion for
the interaction with the external probe should be taken to the same
chiral order as the maximum order used in the chiral expansion of the
$NN$ system Hamiltonian.

An alternative to this approach is the Q-counting scheme pioneered by
Kaplan, Savage, and Wise~\cite{Ka98B,Ka98A}.  In this approach
power-counting for S-matrix elements is explicit, with all quantities
being written as an expansion in the ratio of $\sqrt{M B}$, $p$, and
$m_\pi$ to some underlying scale $\Lambda$. In practice, the deuteron
(or the quasi-bound state in the ${}^1S_0$ channel) is generated by
iterating a non-derivative contact interaction to all orders.  This
builds into the theory an explicit low-energy pole; all other
interactions (including one-pion exchange) are then treated
perturbatively, with a power-counting for the short-distance physics
determined by the renormalization group flows of the operators. A
number of observables in the two-nucleon system have been calculated
using Q-counting, and it appears to work well for many observables at
momenta $p \leq m_\pi$.  However, as yet, no observable has been
calculated where this Q-counting approach with explicit pions works
demonstrably better than a similar Q-counting developed for the
pionless effective theory.  This does raise the question of how well
Q-counting actually incorporates the pionic physics of the $NN$
system.

In this paper we adopt Weinberg's power-counting, and make what are
essentially chiral perturbation theory expansions of the $NN$
potential and the two-particle-irreducible kernel for
electron-deuteron scattering.  Our goal is to examine the physics of
deuteron electromagnetic form factors within this approach.  Our
nucleon-nucleon potential has a structure determined from an effective
field theory expansion. Higher-derivative nucleon-nucleon interactions
are suppressed by appropriate powers of the characteristic
short-distance scale $\Lambda$.  Accordingly we follow Lepage's
usage~\cite{Le99} and refer to this scheme as $\Lambda$-counting.  We
will demonstrate that this approach to effective field theory can be
used to generate deuteron wave functions which are in accord with the
wave functions of much more sophisticated, albeit more
phenomenological, nucleon-nucleon potential models, for all radii $r$
larger than about $1/m_\pi$. The difference between these two
different types of wave function then occurs almost entirely at short
distances, i.e. distances less than $1/m_\pi$.  So, if the central
tenet of effective field theory is indeed correct in the deuteron
system these simpler wave functions should be adequate for the
calculation of low-energy observables. To put this the other way
around, if our wave functions yield predictions for low-energy
quantities which differ significantly from those of potential models
then this represents an unexpected sensitivity to short-distance
dynamics. Such a sensitivity is unexpected precisely because it is
contrary to a very basic belief which underpins the success of
effective field theory.

Of course, these wave functions are not physical observables.
Nevertheless, as will be discussed below, the asymptotic tails of the
wave functions, can, in principle, be obtained from
experiment~\cite{LL77}.  Physical observables dominated by such tails
should be well described in our approach.  More generally, the EFT
framework used here gives the nonrelativistic impulse approximation as
the leading nontrivial~order for many observables. As is well known,
in this approximation for electron-deuteron scattering the wave
functions for particular deuteron magnetic sub-states are probed.
Therefore we can test the ``central tenet'' by computing the three
electromagnetic form factors of the deuteron using our
effective-field-theory-motivated wave functions and comparing them to
results from potential model wave functions.  This allows us to
examine how large the momentum transfer $Q$ must be before one starts
probing the physics at scale $\Lambda_\chi$ which was not explicitly
included in the development of the deuteron wave function.

Thus, at the heart of this paper is a calculation of electron-deuteron
scattering in effective field theory. A calculation of this process in
Q-counting recently appeared~\cite{Ka98C}, but until now
$\Lambda$-counting has not been applied to this reaction except at
zero momentum-transfer~\cite{Or96,Pa98}. Perhaps this is because the
results are not very surprising! Indeed, we shall demonstrate that
$\Lambda$-counting provides a systematic justification for many of the
features observed in potential model calculations of the deuteron
electromagnetic form factors: the importance of corrections associated
with nucleon structure, the unimportance of relativistic corrections
at low $Q^2$, and the strong suppression of meson-exchange current
mechanisms at these momentum transfers. This approach leads to simple
power-counting rules to estimate errors for the form factors $F_C$,
$F_Q$, and $F_M$, and we will calculate these quantities up to the
next-to-leading nontrivial~order in our expansion.  Moreover, for
certain long-distance observables---such as $\langle r^{2n}
\rangle$---the errors can be estimated even more accurately. The
reason that this is possible is because such moments are dominated by
the tails of the wave function which, as noted above, are physical
quantities.

These arguments show that the deuteron electromagnetic form factors
can be concretely related to the deuteron wave function, provided that
one works only at low orders in the $\Lambda$-expansion.  So, we would
suggest that these form factors at momenta $Q \sim 2 m_\pi$ are a good
place to test the efficacy of the competing schemes' reproduction of the
physics of the pion range inside the deuterium nucleus. In particular,
this may shed some light on whether one-pion exchange can be treated
perturbatively in the two-nucleon system. As mentioned above, it is a
consequence of Q-counting that one-pion exchange should be
sub-leading, with the leading effect provided by the
unnaturally-enhanced contact operators that yield a bound deuteron.
Cohen and Hansen have shown that such an expansion for the $NN$
amplitude can be obtained from {\it any} potential $V_{\rm OPE} +
V_{\rm short}$, provided that $m_\pi R \ll 1$, where $R$ is the range
of the short-distance potential $V_{\rm short}$~\cite{CH98A}. While
this shows that one-pion exchange can indeed be treated perturbatively
in a particular formal limit it remains unclear whether $m_\pi$ is
small enough in the physical world to make the results thereby
obtained useful~\cite{Ge98C}.

Our paper is structured as follows. In Sec.~2 we review the basic
physics of nucleon-nucleon potential models, provide a more detailed
introduction to effective field theory than the thumbnail sketch given
here, and outline our particular approach to the calculation of
deuteron wave functions and electromagnetic properties. Section 3
contains the details of our calculation of this wave function, along
with some specific error estimates for quantities like $\langle r^{2n}
\rangle$. Section 4 gives a description of the expansion that
$\Lambda$-counting provides for the kernel of the electron-deuteron
scattering reaction.  In Sec.~5 we present our results from this
calculation. These include results for static deuteron properties,
which are shown to have errors in accord with our systematic error
estimates, and results for $F_C$, $F_Q$ and $F_M$. Note that in the
latter cases we do not compare with experimental data, since potential
model calculations are already very successful for these observables.
Rather, by comparing the predictions for $F_C$, $F_Q$ and $F_M$ with
those of potential models we see that the ``central tenet'' of
effective field theory is alive and well in the deuteron.  Finally our
results are summarized and our conclusions discussed in
Section~\ref{sec-conclusion}.

\section{Potential models, effective field theory, and error estimates}

Traditionally, low-energy phenomena in light nuclei have been
described using the Schr\"odinger equation with a model for the $NN$
potential. In practice, all sensible potentials include the
one-pion exchange~ potential (OPEP) as their long-range tail.  The
inclusion of OPEP is required on theoretical grounds, as it uniquely
describes the longest-range part of the strong interaction. This can
also be seen experimentally, since OPEP is essential in the
description of the higher partial waves in $NN$ scattering. On the
other hand, apart from their common one-pion exchange tails, different
$NN$ potential models look rather dissimilar.  Some include a set of
single (or, in some cases, double) boson-exchanges to generate the
short-distance piece of the interaction, as in the
Nijmegen~\cite{Na78,St94,St97} or the original Bonn
potentials~\cite{Ma87,Ma89}.  Others have a more phenomenological
short-distance structure, e.g. the Reid potential~\cite{St94,Re68},
the Paris potential~\cite{Co73,La80}, or the Argonne
potentials~\cite{Wi95,Wi84}~\footnote{The CD-Bonn
  potential~\cite{Ma96} nominally involves single-meson exchanges at
  short distances, but the mass of the scalar-isoscalar meson
  varies from partial wave to partial wave.  Therefore in classifying
  OBE potentials as we have done here it is not entirely clear which
  class to place it in.}. In all cases the short-distance potential
must be rich enough to describe the existing data.  Once a particular
form is chosen the Schr\"odinger equation is solved and the free
parameters are fit to the $NN$ scattering data and certain deuteron
bound state~properties. 

This is now an extremely sophisticated
enterprise.  Potentials exist which fit the $np$ and $pp$ scattering
data up to laboratory energies of 350 MeV with a $\chi^2$ per degree
of freedom close to one~\cite{St94,Wi95,Ma96}. Such models also do a very
good job in describing a wide array of low-energy data obtained using
electroweak probes. To calculate the response of the nucleus to these
probes, single-nucleon currents and potential-model wave functions are
often employed.  In some cases such a construction does not lead to a
conserved electromagnetic current, and so additional ``model
independent'' two-body currents are added to restore current
conservation (see, for instance~\cite{Ri84}).  Conversely, transverse
currents are not constrained by current conservation and so are always
``model dependent''. The forms chosen reflect dynamical assumptions
about the important physical mechanisms, e.g. that pionic,
$\rho$-meson, $\rho \pi \gamma$, $\omega \pi \gamma$, $\Delta \pi
\gamma$, and $\Delta \rho \gamma$ meson-exchange currents dominate the
transverse currents. For a thorough review of this approach and its
application in few-nucleon systems see Ref.~\cite{CS98}.

Thus, at the very least, potential models represent an extremely
efficient way to describe the data. On the other hand, one difficulty
of the approach is that it is hard to know what is learned about the
dynamics from the success of the description. The fact that a variety
of potential models with different short-distance dynamics all do well
in describing the data suggests that a good description of the data
does not necessarily indicate identification of the underlying
short-distance processes.  However, from an effective-field-theory
viewpoint this is entirely to be expected.  Indeed, the central
tenet of effective field theory is that it is not possible to resolve
details of short-distance physics from low-momentum observables. Thus
in an EFT framework, attempts to get the ``right'' short-distance $NN$
potential using low-momentum data are futile.  This suggests that the
limitation be viewed as an opportunity. If details of short-distance
physics did strongly affect low-energy observables it is very unlikely we would
ever make any good predictions at low energy.  After all, there are an
infinite number of choices for the short-distance physics and we are
unlikely to stumble on the right one. Potential models have predictive
power precisely because the details of the short-distance physics do
not matter greatly.  Thus, the very success of a variety of $NN$
potential models suggests that the $NN$ system is well suited to an
EFT treatment.

Having said this, there are a number of different views on how to
implement effective field theory ideas in nuclear physics.  That
originally proposed by Weinberg~\cite{We90,We91} and employed by
Lepage in Refs.~\cite{Le99,Le97} allows one to think straightforwardly
about the connection of EFT and potential models.  As discussed in the
Introduction, in this approach, known as ``$\Lambda$-counting'', the
focus is on power counting at the level of the Hamiltonian in the $NN$
piece of the hadronic Hilbert space. The effective Lagrangian consists
of $NN$ contact interactions together with the standard heavy baryon
chiral perturbation theory Lagrangian:
\begin{equation}
{\cal L}={\cal L}_{\rm HB \chi \rm PT} + {\cal L}_{NN}.
\end{equation}
At lowest order ${\cal L}_{\rm HB \chi PT}$ may be written in a form
which explicitly shows that at this order the interaction terms
involving zero- and one-pion field take the form of a standard
nonrelativistic Lagrangian for the interactions of pions, nucleons
and photons~\cite{Ka98C}:
\begin{eqnarray}
  && {\cal L}_{\rm HB \chi PT}=\nonumber\\ && \quad N^\dagger \left(i
    D_0 + \frac{{\bf D}^2}{2 M} \right) N + \frac{g_A}{2 f_\pi}
  N^\dagger \, {\bf \sigma} \cdot ({\bf D} (\vec{\tau} \cdot
  \vec{\pi})) \, N + \frac{1}{2} [(D_\mu \vec{\pi}^\dagger D^\mu
  \vec{\pi}) - m_\pi^2 (\vec{\pi}^\dagger \vec{\pi})] +
  \frac{1}{2}({\bf E}^2 - {\bf B}^2), \nonumber\\
\label{eq:NRL}
\end{eqnarray}
where $f_\pi=93$ MeV, and $N$ and $\vec{\pi}$ are the nucleon and pion
isodoublet and isotriplet.  Chiral symmetry is not manifest here, as
the Lagrangian written in terms of the objects which transform
appropriately under $SU(2)_L \times SU(2)_R$ has been expanded out in
terms of the more familiar $\vec{\pi}$ to yield Eq.~(\ref{eq:NRL}).
${\bf E}$ and ${\bf B}$ are the electric and magnetic fields of the
photon, and $D_\mu$ is the usual covariant derivative, which ensures
that $D_\mu N$ transforms in the same way as $N$ under the group
$U(1)_{em}$:
\begin{equation}
D_\mu=\partial_\mu + i e Q_{em} A_\mu + \ldots,
\end{equation}
with $Q_{em}$ the isospin-space matrix:
\begin{equation}
Q_{em}=\left( \begin{array}{cc}
                  1 & 0 \\
                  0 & 0 \\
              \end{array}
\right),
\end{equation}
and $A_\mu$ the photon field. The ellipses indicate that additional
terms in $D_\mu$ involving one or more pion fields are needed if it
is to also be a covariant derivative under $SU(2)_L \times SU(2)_R$.
These terms do not play a role in any of our calculations.
Meanwhile the S-wave piece of the Lagrangian containing the
four-nucleon contact terms is:
\begin{equation}
{\cal L}_{NN}=-\frac{1}{2} C_0 (N^\dagger N)^2 
- \frac{1}{2} C_2 [(N^\dagger N)(N^\dagger {\bf D}^2 N)
+ (N^\dagger N)({\bf D}^2 N^\dagger N)] + \ldots.
\end{equation}

This Lagrangian can then be used to generate an $NN$ potential, $V$ as
follows. In principle, $V$ contains all two-particle-irreducible $NN
\rightarrow NN$ graphs. As shown by Weinberg and Ordo\~n\'ez et
al.~\cite{We90,We91,Or96}, each of these graphs is of a given order in
the chiral expansion, and more complicated graphs occur at higher
order. The other issue here is how to count the contact interactions
$C_{2n}$.  In the framework of ``$\Lambda$-counting'' for $V$, we
simply use naive dimensional analysis in this nonrelativistic field
theory to estimate:
\begin{equation}
C_{2n} \sim \frac{1}{M \Lambda^{2n+1}}.
\label{eq:C2nsize}
\end{equation}
It follows that at leading order in $\Lambda$-counting, the $NN$
potential has two pieces~\cite{We91}:

\begin{equation}
  V({\bf q})=-\frac{g_A^2}{4 f_\pi^2} \, \frac{{\bf \sigma}_1 \cdot {\bf
      q} \, {\bf \sigma}_2 \cdot {\bf q}} {{\bf q}^2 + m_\pi^2} \, 
  (\vec{\tau}_1 \cdot \vec{\tau}_2) + C_0,
\label{eq:VNN0}
\end{equation}
where $C_0$ is a contact interaction which is a constant in momentum
space (or equivalently, a three-dimensional delta-function in
coordinate space). This contact operator gives the short-distance
piece of the $NN$ interaction, while one-pion exchange gives the
long-range tail.  Indeed, at any chiral order an analogous separation
into a long- and short-distance $NN$ EFT potential can be made.  At
low orders in the expansion the short-distance part has
straightforward $\Lambda$-counting, with the higher-derivative
operators always suppressed by powers of $p/\Lambda$.  Meanwhile, the
long-distance piece is given by one-pion exchange plus irreducible
two-pion exchange plus three-pion exchange plus etc.
etc.~\footnote{In practice, three-pion exchange is of a range
  comparable to that of the ``short-distance'' potential, but, in the
  chiral limit it is clear that such a separation into long and
  short-distance pieces could be made unambiguously.} The dimension of
the operators which these graphs produce in the $NN$ Hamiltonian turns
out to increase by two for each additional pion
loop~\cite{We90,We91,Or96}. Consequently, these irreducible multiple
pion exchanges are also suppressed by powers of $m_\pi/\Lambda$ and
$p/\Lambda$. Both this long-range part of $V$ and the short-range part
coming from the contact operators should then be truncated at the same
order in the expansion in $p$ and $m_\pi$ over $\Lambda$.

The resulting potential is then used nonperturbatively~ in a
Schr\"odinger equation.  Of course, this procedure will generate
divergences unless the contact terms are regulated.  This is achieved
by taking some (essentially arbitrary) functional form parameterized
by a regulator mass to serve as a regulated coordinate space
three-dimensional delta function.  Practical implementation of the
renormalization program is then quite simple: with the regulator mass
fixed, one tunes the coefficients so as to reproduce some specified
low-energy observables~\cite{Le97,Sc97}. In Ref.~\cite{Or96} such a
fit to both $NN$ phase shifts up to laboratory energies of 100 MeV and
deuteron properties was done with the next-to-next-to-next-to-leading
order (NNNLO) $\Lambda$-counting interaction. Three different
regulator parameters and a Gaussian regulator were employed to render
finite the divergences which would otherwise have appeared.

With renormalizable theories, the next step would be to take the
regulator mass to infinity.  In Refs.~\cite{Be97B,Ge98B,Le97,Le90}, it is
argued that this should not be done for these effective theories.
Rather, the regulator mass is to be set at, or near, the scale of the
short-distance physics, $\Lambda$.  One advantage in doing this is
that one then expects that the bare coefficients in the expansion will
be of the size given by Eq.~(\ref{eq:C2nsize}), whereas if regulator
scales were chosen to be very different from $\Lambda$ the values of
the bare coefficients could be quite different from this
naive-dimensional-analysis estimate.  Moreover, by keeping the
regulator finite and of order $\Lambda$ one avoids the difficulties
discussed in Refs.~\cite{PC97,Ph97}.

This particular treatment of short-distance physics in EFT is actually
not very different from that employed in some of the traditional
potential models described above. Any functional form given by a
potential model with characteristic scale $\Lambda$ and $n$ parameters
could be rewritten as the sum of $n$ terms with each term given by
derivatives of some ``regulated delta-function''. Thus, the gain
achieved in using EFT as compared to traditional potential models is
really the ability to treat the long-distance physics and the current
operators in a manner consistent with each other and with the
short-distance physics by working to a given order in
$\Lambda$-counting in the system's Hamiltonian. In particular
$\Lambda$-counting constrains the number and form of current operators
allowable at a given order and mandates the type of pion-loop and
pion exchange~ mechanisms which should be included in the calculation.
Such formal consistency is very satisfying.  It also confers the
practical advantage that errors can be estimated in a reliable way.
The key point is that the characteristic size of the physics neglected
at each stage of the calculation is known.  In particular, in this
work we use the EFT approach to analyze uncertainties in calculations
of electromagnetic properties of the deuteron.

The errors which arise include uncertainties in our wave functions and
uncertainties in the current operators---both of which are due to
truncations in the $\Lambda$-expansion. In discussing these errors we
first focus on the tail of the bound state~wave function, which, while
not directly observable, can be obtained indirectly from scattering
data. This can be seen simply in a spectral representation of the
general off-shell $NN$ t-matrix.  The t-matrix diverges as one
approaches a bound state pole in energy. Then, if there is a bound
state $|\phi \rangle$ at $E=-B$ the residue there is directly related
to the bound state~density-matrix via:

\begin{equation}
  \lim_{E \rightarrow - B} \, (E + B) \, T(E) \, = \, (B + H_0) \, |
  \phi \rangle \langle \phi| \, (B + H_0),
\end{equation}
where $H_0$ is the free Hamiltonian. Now, consider both sides of this
equation in coordinate space. In both the initial and the final state
take the relative distance between the two nucleons to be much larger
than the range of the interaction. For positive energies the left-hand
side can be expressed in terms of the measured $NN$ phase shifts. This
gives the coordinate-space tail of the scattering wave function. An
extrapolation to the bound state~pole at $E=-B$ may then be made to
obtain the tail of the bound state~wave function.  Thus, for systems
with a shallow bound state, accurate scattering data determines the
asymptotic tail of the bound state~wave function in a
model-independent way.

Since our interest here is in long-distance phenomena in the deuteron,
it seems sensible to choose our fitting procedure to ensure that the
asymptotic deuteron wave function is correct. If some other
fitting-procedure were employed, e.g. fitting the coefficients in a
Taylor expansion of the $T$ matrix or $1/T$, additional errors will be
introduced, since the tail of the deuteron wave function would acquire
some errors as part of the extrapolation from finite $p$ to $p=i
\sqrt{MB}$. These errors are formally higher-order in the power
counting. Indeed, they are very small in $\Lambda$-counting, as
demonstrated in a calculation without explicit pions by Park et al. in
Ref.~\cite{Pa97}. There the ${}^3S_1$ scattering length and effective
range were fitted and $A_S$ was then reproduced to about 1\%
accuracy. Since we are free to choose among the different possible
fitting procedures, and these are all equivalent to the order we work
here, we fit to the experimental values for the deuteron binding
energy $B$, and for the asymptotic S and D-state normalizations $A_S$
and $A_D$.  This has the merit that the longest-distance parts of the wave
function is described perfectly, within the experimental errors. In
particular, the deuteron pole appears at the correct location in the
complex energy plane, and has the correct residue.

Proceeding in this way we integrate this essentially-exact asymptotic
form from $r=\infty$ inward. Errors become large when we reach
one-pion ranges if only the free Schr\"odinger equation is used to do
this integration. If one-pion exchange is added to the differential
equation used in integrating inwards the resulting wave functions
are accurate down to distances $r \sim 1/m_\pi$.  Adding
two-pion exchange would further increase the range over which the wave
functions are accurate. Moreover, as we shall show in
Section~\ref{sec-TPE} the chiral suppression of two-pion exchange
noted above means that the effects of the two-pion exchange~ potential
will be small even at ranges where they might be thought to be
important.  Similarly, relativistic corrections to both the nucleon
motion and the one-pion exchange~ potential are suppressed by powers of
$p/M$, where $p$ is the typical momentum inside the bound state. It is
therefore quite easy to demonstrate that the wave functions employed
are accurate in the region $r > 1/m_\pi$ up to corrections of
relative order $(P/\Lambda)^2$, where $P$ could be either the typical
nuclear momentum, $p$, or $m_\pi$, and $\Lambda$ is $\Lambda_\chi$ or
$M$.

Of course, in pursuing the wave function to ever smaller $r$
ultimately one reaches a distance $R$ which is so short that the
chiral expansion for the potential fails, and some physics beyond the
range of the effective theory starts to play a role.  At this point we
employ an arbitrary regulator, motivated by the regulated
delta-function of $\Lambda$-counting, to ensure that the wave function
has the correct behavior as $r \rightarrow 0$. The way in which the
deuteron wave function is obtained in this ``integrating in'' picture
is very similar to that employed for scattering wave functions in the
Nijmegen phase-shift analyses~\cite{St93,Re99}.

For probes of the deuteron which probe the scale $R$ the error
analysis employed in this paper breaks down completely, since the wave
function for $r \leq R$ is largely fallacious. On the other hand, it
is one of the central conclusions of this paper that electron-deuteron
scattering is {\it not} such a probe, at least provided that $Q \leq 700$
MeV, and reasonable values for $R$ are chosen.

Thus far we have only spoken about errors in the wave function. Of
course, there are corresponding errors in the current. In this work we
employ only the next-to-leading (NLO) order chiral-perturbation theory
deuteronic current. That current is derived in Sec.~\ref{sec-kernel}.
It represents two particles of charge $e$ and zero and anomalous
magnetic moments $\mu_p$ and $\mu_n$. No terms corresponding to the
nucleon's charge or magnetic radius appear in the current at this
order. Such terms enter as a correction at the next order in the
expansion. Other corrections to the one-nucleon current also occur one
order beyond that considered here. As for two-body currents, as
explained in Sec.~\ref{sec-kernel}, in the chiral expansion {\it all}
such effects are suppressed by {\it at least} two powers of
$P/\Lambda$ relative to the leading-order operator.

If the probe in question does not directly probe the wave function
for $r \leq R$, we can write the matrix element of some current operator
as:

\begin{equation}
  \langle \psi | O | \psi \rangle=(\langle \psi^{(0)} | + \langle
  \delta \psi|) (O^{(0)} + \delta O) (|\psi^{(0)} \rangle + |\delta
  \psi\rangle),
\end{equation}
where the operator $O_0$ and the wave functions $|\psi^{(0)} \rangle$
are the ones employed in this calculation, and the arguments of the
following two sections will show that the relative size of both
$\delta O$ and $|\delta \psi \rangle$ is $P^2/\Lambda^2$.  This means
that the electron-deuteron observables calculated here should be
accurate up to corrections of relative order $P^2/\Lambda^2$, provided
that $Q$ is not so large as to probe the short-distance piece of the
deuteron wave function. This justifies the nonrelativistic
impulse approximation as both the leading and next-to-leading term in
this effective field theory expansion of deuteron electromagnetic
properties. Furthermore, we will see in Section~\ref{sec-kernel} that
the pieces of $\delta O$ which appear at relative order
$P^2/\Lambda^2$ and $P^3/\Lambda^3$ are those which are known to be
important in potential-model calculations of deuteron electromagnetic
properties.

\section{Wave functions}

\label{sec-wavefunction}

Here we discuss the wave functions which we use to
calculate the deuteron electromagnetic form factors.  One thing we aim
to test in this work is at what virtual photon momentum the details of
the short-range deuteron structure affect the electromagnetic
observables $F_C$, $F_Q$, and $F_M$.  To this end we think of the wave
functions as being generated by integrating the Schr\"odinger equation
in from infinity, given a certain asymptotic tail for the wave
function. The parameters of this asymptotic tail can be determined
experimentally, at least in the sense of dispersion relations. In the
deuteron they are the deuteron binding energy $B$, the asymptotic
S-state normalization, $A_S$, and the asymptotic D/S ratio, $\eta$.
(These were also the parameters chosen in Ref.~\cite{Pa98}.)  If no
other interaction were included and the asymptotic wave functions were
taken to be true all the way into the origin this would be the
deuteron of the so-called zero-range approximation, but with the
experimental value for $A_S$ (see, for instance~\cite{Bl88}).  These
zero-range wave functions can be improved upon at one-pion ranges if
we do not take the asymptotic form all the way in to $r=0$, but
instead include one-pion exchange in the Schr\"odinger equation when
integrating in from $r=\infty$. In this way two different kinds of
wave functions, both of which have the incorrect behavior as $r
\rightarrow 0$ can be generated, with no assumptions made about the
short-distance piece of the $NN$ interaction: ``zero-range'' wave
functions, in which the asymptotic form persists to $r=0$, and wave
functions which are ``integrated in'' using one-pion exchange to all
orders. The details of this procedure are discussed in
Section~\ref{sec-wfdetails}.

In order to render our calculations finite we must also
introduce a short-distance regulator. This procedure is discussed in
Sec.~\ref{sec-shortdistance}.  There, we choose a square well for
this purpose, and determine the wave functions for different values of
the square-well radius, $R$. Note that all these wave functions have
exactly the same structure outside of the square well, since we use
the method of Sec.~\ref{sec-wfdetails} to construct their tails.

Introducing this arbitrary, and undoubtedly incorrect, short-distance
potential leads immediately to the question of whether this procedure
introduces uncontrollable errors for observables.  In
Sec.~\ref{sec-errors} we show how, in this potential model, the
error introduced by our lack of knowledge of the short-distance piece
of the wave function can be estimated for the moments $\langle r^{2n}
\rangle$.

In Secs.~\ref{sec-TPE} and \ref{sec-relativisticcorrections}
we discuss how to systematically estimate the errors which arise from
the inaccuracy of our wave functions {\it outside} $r=R$. Namely, we
ask how the introduction of two-pion exchange, the $\pi NN$ form
factor, and relativistic effects in the deuteron wave function would
affect the error bounds derived in Sec.~\ref{sec-errors} for $\langle
r^{2n} \rangle$.

Not surprisingly, the size of these errors depends crucially on the
radius $R$ at which the short-distance potential starts, so we
provide estimates for $R$ in Section~\ref{sec-estimates}.

\subsection{Calculating the wave function by ``integrating in''}

\label{sec-wfdetails}

In this section we explain how the deuteron wave functions
\begin{equation}
|M \, \, {\bf P} \rangle
\end{equation}
are calculated.

As explained above, we work in an effective field theory in which
nucleons are treated nonrelativistically and relativistic effects are
included as a perturbation. Below we will quantify the effect of
relativistic corrections. But, for the moment we make the usual
nonrelativistic~ decomposition of the wave function $|M \, \, {\bf P}
\rangle$:
\begin{equation}
  |M \, \, {\bf P} \rangle=\int \frac{d^3 p}{(2 \pi)^3} |{\bf P} \, \,
  {\bf p} \rangle \sum_{L S} \sum_{m_S m_L} \tilde{u}_L(p) \, \, (L \, \, m_L
  \, \, S \, \, m_S|J \, \, M) \langle \hat{p}|L \, m_L \rangle \, \,
  \, |S \, m_S \rangle,
\end{equation}
where $\langle \hat{p}|L \, m_L \rangle$ are the usual spherical
harmonics, $(L \, \, m_L \, \, S \, \, m_S|J \, \, M)$ are the
Clebsch-Gordon coefficients, $\tilde{u}_L$ is the radial wave function
corresponding to orbital angular momentum $L$, and the spin wave
function $|S \, \, m_S \rangle$ can be expressed in terms of the
single-nucleon spins via:
\begin{equation}
  |S \, \, m_S \rangle=\sum_{m_1 m_2} (1/2 \, \, m_1 \, \, 1/2 \, \,
  m_2|S \, \, m_S) |1/2 \, \, m_1 \rangle |1/2 \, \, m_2 \rangle.
\end{equation}

The wave functions $\tilde{u}_L(p)$ are the spherical Bessel transforms of the
radial wave functions $u_L(r)$,
\begin{equation}
\tilde{u}_L (p)=i^L 4 \pi \int dr \, j_L(pr) \, r u_L(r),
\end{equation}
where $j_L$ is a spherical Bessel function.

In the case of the deuteron the spin and total angular momentum of the
state are $S=J=1$, and so the states $L=0$ and $L=2$ can contribute to
the wave function.  Following convention we define $u(r)=u_0(r)$ and
$w(r)=u_2(r)$.

Suppose that the asymptotic wave functions $u_0(r)$ and $u_2(r)$ are
known. Then we wish to solve the radial Schr\"odinger equation:
\begin{equation}
-\frac{d^2 u_l}{d r^2} + \frac{l(l+1)}{r^2} u_l(r) + 
\sum_{l'=0,2} M V_{l l'}(r) u_{l'}(r)=-\gamma^2 u_l(r),
\label{eq:radSE}
\end{equation}
where $\gamma^2=M B$, and we have allowed for the possibility that $V$
is a non-central potential which mixes states of different orbital
angular momentum. This equation is to be solved subject to the
boundary conditions:
\begin{eqnarray}
  u(r) &\longrightarrow& A_S e^{-\gamma r} \mbox{ as } r
  \longrightarrow \infty, \label{eq:uas}\\
  w(r) &\longrightarrow& A_D \left(1 + \frac{3}{\gamma r} +
    \frac{3}{(\gamma r)^2} \right) e^{-\gamma r} \mbox{ as } r
  \longrightarrow \infty; \label{eq:was}
\end{eqnarray}
where $\gamma$ and the asymptotic normalizations $A_S$ and $A_D$ 
are given quantities.

In the pionless theory $V_{l l'}(r)=0$ and so these asymptotic forms
persist all the way into $r=0$. We define:
\begin{eqnarray}
u^{(0)}(r)&=&A_S e^{-\gamma r},\\
w^{(0)}(r)&=&A_D \left(1 + \frac{3}{\gamma r} +
    \frac{3}{(\gamma r)^2} \right) e^{-\gamma r};
\end{eqnarray}
as these so-called ``zero-range'' solutions.

In the theory with pions we have the following angular-momentum decomposition
for the one-pion exchange~ potential in the deuteron channel:
\begin{eqnarray}
V_{00}(r)&=&V_C(r) \label{eq:OPE1}\\
V_{02}(r)&=&V_{20}(r)=2 \sqrt{2} V_T(r)\\
V_{22}(r)&=&V_C(r) - 2 V_T(r),
\end{eqnarray}
where the central and tensor pieces of the potential are, respectively:
\begin{eqnarray}
V_C(r)&=&-m_\pi f_{\pi NN}^2 \frac{e^{-m_\pi r}}{m_\pi r}, \label{eq:VC} \\
V_T(r)&=&-m_\pi f_{\pi NN}^2 \frac{e^{-m_\pi r}}{m_\pi r}
\left(1 + \frac{3}{m_\pi r} + \frac{3}{(m_\pi r)^2} \right). \label{eq:OPE5}
\end{eqnarray}
To leading order in the chiral expansion the $\pi NN$ coupling is given by
the Goldberger-Treiman relation:
\begin{equation}
f_{\pi NN}^2=\frac{g_A^2 m_\pi^2}{16 \pi f_\pi^2},
\end{equation}
although we will not be using this exact value in our work.

We label the wave functions $u$ and $w$ found by solving
Eq.~(\ref{eq:radSE}) with this one-pion exchange~ potential as
$u_\pi(r)$ and $w_\pi(r)$. Similar solutions were obtained by
Klarsfeld {\it et al.},~ in the early 1980s~\cite{Kl81,Kl84}, and similar ideas
were used to generate a D-wave wave function given an S-wave wave
function, $u$, by Ericson and Rosa-Clot~\cite{ERC83,Er84,ERC85}.

\subsection{Short-distance regulator}

\label{sec-shortdistance}

Note that we have dropped the delta-function piece of the central
potential in Eq.~(\ref{eq:VC}), since we will never integrate all the
way into $r=0$.  Instead, for radii $r$ less than some specified
distance $R$ we replace one-pion exchange by a central short-distance
potential $V_S(r)$. This short-distance interaction subsumes all the
details of the physics of the nucleon-nucleon system at short range.
It is designed to ensure the correct $r \rightarrow 0$ behavior of $u$
and $w$, namely:

\begin{equation}
u(r) \sim r; \quad w(r) \sim r^3 \, \mbox{ as } r \rightarrow 0.
\label{eq:rgoestozero}
\end{equation}
Without the presence of $V_S(r)$ $u$ and $w$ do not have the
appropriate $r \rightarrow 0$ limits, and divergences in physical
quantities may result.  Once these divergences are removed
quantities which are dominated by long-distance physics should not be
sensitive to the details of the short-distance potential. 

Thus we are encouraged to choose a form for the short-distance
potential which allows us to compute the wave function for $r < R$
analytically.  Hence we choose the sum of a square-well and a surface
delta-function. The potentials $V_{l l'}(r)$ retain the form
(\ref{eq:OPE1})--(\ref{eq:OPE5}) for $r > R$ in the theory with pions
and are zero in the theory without pions. Then, in the S-wave, for $r
< R$ we have the structure:

\begin{equation}
V_{00}(r)=V_0 \theta (R - r) + V_1 \delta(r - R),
\label{eq:VS}
\end{equation}
while in the D-wave:

\begin{equation}
V_{22}(r)=V_2 \theta (R - r).
\end{equation}
In this ``integrating in'' approach the radius $R$ is given, and the
coefficients $V_0$, $V_1$, and $V_2$ may then be derived from the
conditions of continuity of the wave function and its first derivative
at $r=R$, as well as the condition that the wave function has unit
norm. This is equivalent to fitting $V_0$, $V_1$, and $V_2$ to $B$,
$A_S$, and $\eta$. Such an approach is similar to that employed in
Ref.~\cite{Pa98}. The main difference is that our short-distance
potential is constructed in coordinate space, and has a different
angular-momentum structure than that of Park {\it et al.}

Specifically, the wave function inside $r=R$ is:
\begin{equation}
u_l(r)=C_l r j_l(\alpha_l r),
\label{eq:uinside}
\end{equation}
where $\alpha_l^2=M (V_l - B)$. Since we have integrated the
Schr\"odinger equation with or without one-pion exchange in to $r=R$
the value of $u_l$ at $r=R$ is known. The coefficients $C_l$ are then
easily seen to be:
\begin{equation}
C_l=\frac{u_l(R)}{R j_l(\alpha_l R)}. 
\end{equation}
The coefficient $V_1$ is set by the requirement that it produce the
appropriate change in the first derivative of $u_0$ at $r=R$.  

Meanwhile, $\alpha_2$ is determined by the requirement that the
logarithmic derivative of $w(r)$ be continuous, and $\alpha_0$ by the
requirement that the wave function be normalized to unity. We may
write:
\begin{equation}
\int_0^R dr \, u^2(r) =1 - \int_R^\infty dr \, u^2(r) - \int_0^\infty 
dr \, w^2(r),
\label{eq:alphaeqn}
\end{equation}
where the quantity on the right-hand side can be calculated
numerically.  Since $u$ and $w$ have the form (\ref{eq:uinside}) for
$r \leq R$ the left-hand side may be calculated analytically and an
equation for $\alpha_0$ obtained and solved. 

Observe that if $V_S$ is set to zero, and the equations
(\ref{eq:radSE}) are solved all the way into the origin, then the wave
function violates unitarity (it has norm greater than one) and does
not obey Eq.~(\ref{eq:rgoestozero}).  Of course, the first problem can
be removed by adding to the probability density from the wave function
for $r > 0$ a piece of negative probability density which contributes
to the overall probability density only at $r=0$. One way to implement
this consistently is using the energy-dependent pseudo-potential
discussed in Ref.~\cite{vK98}. In that case the wave-function
normalization condition is no longer simply $\langle \psi | \psi
\rangle=1$, since there will is a contribution to the normalization
integral from the energy-dependent potential. When this is evaluated
properly it transpires that we still have $F_C(0)=1$, even though the
wave function apparently violates the argument given here and
below. However, if we restrict ourselves to the usual framework of
non-relativistic quantum mechanics, where probability densities are
positive-definite and potentials are energy-independent, we see that
both the theory with pions and the theory with pions integrated out
have a minimum value of $R$ for which Eq.~(\ref{eq:alphaeqn}) has a
solution. If $R$ is made too small the quantity on the right-hand side
becomes negative and so cannot equal the (assumed) positive-definite
quantity on the left-hand side. If one-pion exchange is used in
integrating the wave function in from infinity then the minimum value
of $R$ for which (\ref{eq:alphaeqn}) has solutions for the
short-distance potential (\ref{eq:VS}) is 1.3 fm. Meanwhile, if
one-pion exchange is switched off, the minimum value for which
(\ref{eq:alphaeqn}) has solutions in our calculation is 2.0
fm. Furthermore, if $R < 1.7$ fm and the wave functions $u_0$ and
$w_0$ are used on the right-hand side, then there is no solution for
$R$, no matter what short-distance potential is employed (again,
assuming positive-definiteness of the probability density). Similarly,
if $R < 0.83$ fm and the wave functions $u_\pi$ and $w_\pi$ are used
on the right-hand side then the left-hand side is negative, and so,
again regardless of the short-distance potential employed, no solution
for $R$ exists.  If we assume, as is normally done in non-relativistic
quantum mechanics, that the potentials are energy-independent and 
the wave functions have norm one, then some potential other than
one-pion exchange {\it must} begin to play a role in the dynamics of
the deuteron wave function even at radii $r \sim 1$ fm. Such lower
bounds on the range of the non-one-pion exchange~ piece of the $NN$
potential were derived using similar means by Klarsfeld {\it et al.}
many years ago~\cite{Kl84}.
 
Now, given any $R$ which is large enough to allow a solution to
Eq.~(\ref{eq:alphaeqn}) we can obtain a wave function which has the
following properties:
\begin{itemize}
\item It has the appropriate asymptotic behavior, with values for
$A_S$, $A_D$ and the binding energy taken from experimental data;

\item In the theory with pions it corresponds to the wave function for
the Schr\"odinger equation  solved with one-pion exchange alone 
for all $r$ greater than the specified $R$;

\item It is continuous everywhere;

\item It is normalized to one;

\item It has the appropriate behavior as $r \rightarrow 0$.
\end{itemize}

\begin{figure}[h,t,b,p]
  \vspace{0.5cm} \epsfysize=11 cm
  \centerline{\epsffile{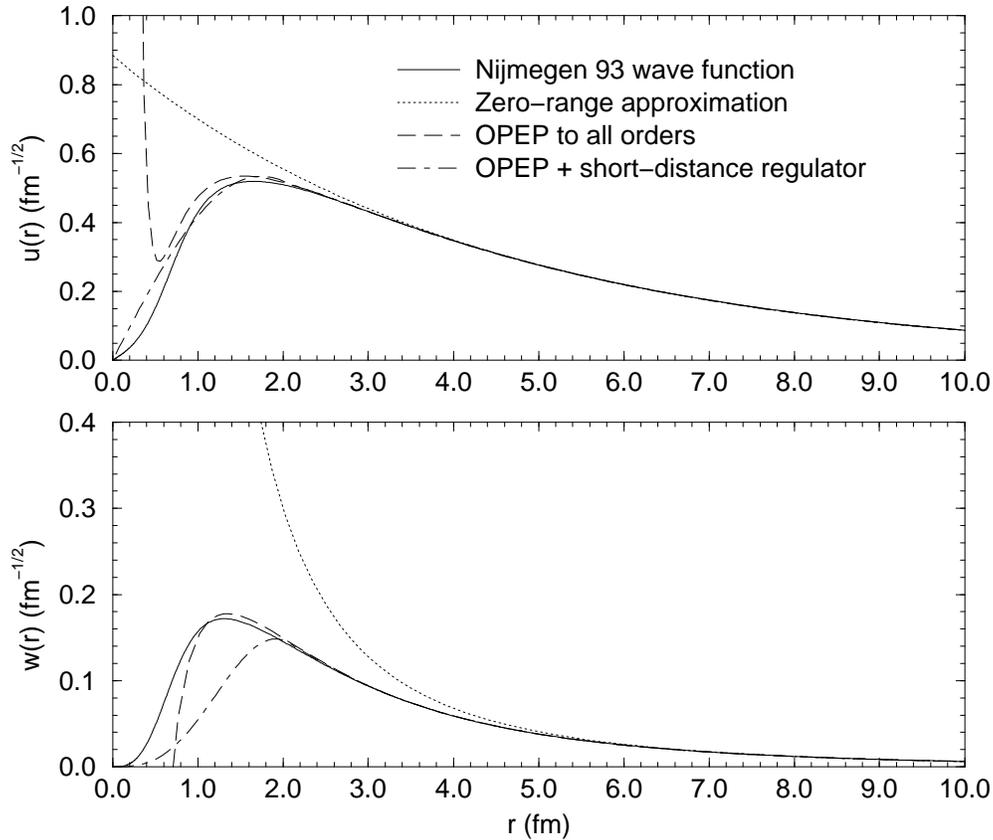}}
  \centerline{\parbox{11cm}{\caption{\label{fig-wfcomparison} The
        radial wave functions of the deuteron, $u(r)$ and $w(r)$, for
        the Nijm93 potential (solid line), as compared to wave
        functions in which the asymptotic form is integrated in to
        $r=0$, both with (dashed) and without (dotted) one-pion
        exchange. The wave function in which the asymptotic form
        persists to $r=0$ is referred to as the ``zero-range''
        solution. The dot-dashed line is the wave function obtained
        when a short-distance regulator is used in the calculation
        with one-pion exchange for $r < 2.0$ fm.  }}}
\end{figure}

The results of this procedure for $R=2.0$ fm are shown for both $u$
and $w$, and compared with $u_\pi$ and $w_\pi$ and $u_0$ and $w_0$, as
well as with the Nijmegen 93 (Nijm93) wave function, in Fig.~\ref{fig-wfcomparison}.
$A_S$, $\eta$, and $B$ were taken to be the central experimental
values~\cite{deS95}:
\begin{eqnarray}
B&=&2.2246 \, {\rm MeV},\\
A_S&=&0.8845 \, {\rm fm}^{-1/2},\\
\eta&=&0.0253.
\end{eqnarray}
Meanwhile, the $\pi NN$ coupling is taken to agree with that of
Ref.~\cite{St94}, so that the wave functions constructed here agree (to
within experimental error bars) with the Nijm93 wave function in the
asymptotic regime and at one-pion ranges:

\begin{equation}
f_{\pi NN}^2=0.075.
\end{equation}

\subsection{Corrections to the wave function for $r \leq R$}

\label{sec-errors}

We will now examine how corrections to the deuteron wave
function inside the range of the short-distance potential affect the
electromagnetic observables under consideration here. We find, not
surprisingly, that the key quantity with which these corrections scale
is $R$.

Consider the observable

\begin{equation}
  \langle r^{2 n} \rangle \equiv \int_0^\infty \, dr \, r^{2 n} \, [u^2(r) +
  w^2(r)],
\end{equation}
where $u$ and $w$ are the ``true'' deuteron wave function~\footnote{Of
  course, the wave function is not observable, and may be changed by a
  unitary transformation (equivalently, by a field redefinition).
  Nevertheless, once we have specified a particular representation
  for the operators in the problem the wave function is a well-defined
  object, and thus we may talk about the ``true'' wave function.}.
Then, provided that the radius $R$ is not too small the wave function
$u_\pi$ will be accurate for $r \geq R$. Thus the error in $\langle
r^{2n} \rangle$, $\delta \langle r^{2n} \rangle$, may be written as:

\begin{equation}
  \delta \langle r^{2n} \rangle=\int_0^R  dr \, r^{2n} \, |u^2(r) +
  w^2(r) - u_\pi^2(r) - w_\pi^2(r)|.
\end{equation}
Now it is simple to bound this quantity from above:

\begin{equation}
  \delta \langle r^{2n} \rangle \leq \int_0^R dr \, r^{2n} \,
  [u^2(r) + w^2(r)].
\end{equation}
Of course, with a ``sensible'' short-distance potential the error
would be much smaller than this, but if the short-distance interaction
were a hard core out to radius $R$, then this would be the error.  To
estimate the right-hand side we now approximate $u$ and $w$ by a
linear and cubic function respectively, yielding:

\begin{equation}
  \delta \langle r^{2n} \rangle \, \lsim \, \int_0^R dr \, r^{2n} \,
  \left[u^2(R)\, \left( \frac{r}{R} \right)^2 + w^2(R) \left( \frac{r}{R}
  \right)^6 \right].
\end{equation}
The quantity $u^2(R)$ may be estimated to be at most of size $2
\gamma$, while $w(R)$ has a size of order $2 \gamma \eta$, with
$\eta=A_D/A_S$ being the asymptotic D/S ratio. Thus the leading
correction comes from the first term and takes the form:

\begin{equation}
  \delta \langle r^{2n} \rangle \, \leq \, \frac{R^{2n}}{2n + 3} \, 2
  \gamma R.
\end{equation}

Naively, we might have expected $\delta \langle r^{2n} \rangle \sim
R^{2n}$, but because the deuteron is a shallow bound state, most of
the weight of the wave function resides in the tail region, where the
dynamics is well known, and so the correction is suppressed by a
further factor of $\gamma R$.  Since the typical size of this quantity
for the shallow bound state is $(2n)!/\gamma^{2n}$ we see that the
fractional error becomes smaller and smaller as $n$ increases, and we
give more weight to the tail

\begin{equation}
  \frac{\delta \langle r^{2n} \rangle}{\langle r^{2n} \rangle}
  =\frac{2}{2n + 3} \, \frac{1}{(2n)!} (\gamma R)^{2n + 1}.
\end{equation}

We have made all these arguments for the effective theory with pions. They 
apply equally well to the one without pions, as long as the radius $R$ is
made appropriately larger. 

We would like to convert these into estimates for the scale of
breakdown of the calculation of $F_C$, $F_Q$ and $F_M$. Consider, for instance,
the expression for $F_C(Q)$ in the nonrelativistic~ impulse approximation:

\begin{equation}
  F_C(Q)=\int_0^\infty dr \, j_0\left(\frac{Qr}{2}\right) \, \, [u^2(r) + w^2(r)].
\end{equation}
This may be expanded in products of powers of $Q^2$ and even moments
of the distribution $u^2(r)$. But that expansion does not converge for $Q
\geq 2 \gamma$. It does, however, give us an estimate of the $Q^2$ at
which our calculation for $F_C$ will become invalid.  Since we already
know the behavior of the errors $\delta \langle r^{2n} \rangle$ we
find

\begin{equation}
  \delta F_C(Q)=\gamma R \sum_{n=1}^{\infty} c_n \left(\frac{Q R}{2}
  \right)^{2n},
\end{equation}
where the $c_n$'ss are dimensionless numbers of order one.  Therefore we
expect that any description of deuteron form factors based on this
approach will cease to be valid when $Q/2 \sim 1/R$. Similar results
hold for the form factors $F_M$ and $F_Q$, although it should be noted
that if the deuteron D-state dynamics is not reproduced with moderate
accuracy these observables will deviate from the experimental data
even at low $Q^2$.

\subsection{The role of two-pion exchange and the $\pi NN$ form factor}

\label{sec-TPE}
 
In the theory with dynamical pions one obvious candidate for $R$ is
the range where two-pion exchange begins to play a significant role in
the dynamics. We see from the recent Nijmegen phase-shift
analysis~\cite{Re99} that this occurs inside $r=1.8$ fm. However,
unlike many short-distance pieces of the $NN$ potential two-pion
exchange is constrained by chiral symmetry. It can be consistently
treated in the framework of chiral perturbation theory (see, for
instance~\cite{Or96,KW97}), and it transpires that the coefficients of
the irreducible two-pion exchange~ kernel are suppressed by two chiral
orders, relative to the coefficients of one-pion exchange. Thus, we
may schematically write the total nucleon-nucleon potential as:
\begin{equation}
V(r)=V_{\rm OPE}(r) + \left(\frac{P}{\Lambda_\chi}\right)^2
V_{\rm TPE}(r) + V_S(r).
\label{eq:VTPE}
\end{equation}
Here $P=m_\pi$ or $p$, where $p$ is the typical momentum inside the
deuteron, which we shall show below is of order $m_\pi$ itself. The
potential $V_{\rm TPE}$ is now nominally the same ``size'' as $V_{\rm
  OPE}$ and $V_S$ acts only in the region $0 \leq r \leq R$.  The
radius $R_\pi$ is defined as the distance at which the
one-pion exchange~ potential wave functions $u_\pi$ and $w_\pi$ begin
to differ significantly from the zero-range wave functions $u_0$ and
$w_0$. We can now identify three regions in the wave function:

\begin{itemize}
\item $r > R_\pi$: Since $\gamma \ll m_\pi$ the wave function is still
  sizable in this region, but it is well described by the asymptotic
  form (\ref{eq:uas})--(\ref{eq:was});

\item $R_{\pi} > r > R$: Here the wave functions $u_\pi$ and $w_\pi$
  are good approximations to the ``true'' wave function.  Two-pion
  exchange can affect the wave function, but its effects are
  controlled by the chirally-small parameter $P/\Lambda_\chi$;

\item $r < R$: Here the wave function is purely the result of a crude
  potential; the only feature in this region which we know to be
  accurate is the $r \rightarrow 0$ behavior (\ref{eq:rgoestozero}).
\end{itemize}

We can assess the value of $R_\pi$ by comparing the zero-range wave
functions $u_0$ and $w_0$ with those $u_\pi$ and $w_\pi$ obtained by
integrating in using the one-pion exchange~ potential.  Due to the
separation of scales $\gamma \ll m_\pi \ll \Lambda_\chi$ there should
be a sizable region where the wave functions in the theory with pions
are good approximations to the ``true'' wave functions and yet differ
significantly from the zero-range wave functions.  The comparison can
be made in Fig.~\ref{fig-wfcomparison} and we conclude that for the
pionless theory 2.5 fm is a reasonable value for the radius at which
$u_\pi$ starts to deviate significantly from $u_0$. Note that for $w$
the distance at which $w_0$ and $w_\pi$ differ significantly is
somewhat larger. In both cases the distances in question are
significantly larger than the naive estimate $R_\pi \sim 1/m_\pi$.

The plots of Fig.~\ref{fig-wfcomparison} confirm that there is
indeed the separation of regions in the wave function required for an
effective field theory treatment of the deuteron to make sense.  Given
such a separation we can identify the correction which two-pion
exchange will make to the observables $\langle r^{2n} \rangle$
discussed above. Firstly, we note that in the region $r > R$ we can
treat $V_{TPE}$ as a perturbation.  Consequently we write:
\begin{equation}
|\psi \rangle=|\psi^{(0)} \rangle + |\delta \psi \rangle,
\end{equation}
where $|\delta \psi \rangle$ is $O(P^2)$ relative to the wave function
$|\psi^{(0)} \rangle$, constructed with $u_\pi$ and $w_\pi$, that was
used above in estimating the errors $\delta \langle r^{2n} \rangle$. 

Next we note that the two-pion exchange correction to $\langle r^{2n}
\rangle$ can only come from the region  $R_\pi > r > R$. Thus, 
if we write the $u$ and $w$ from $|\delta \psi \rangle$ as $\delta u$ 
and $\delta w$ we have:
\begin{equation}
\langle \psi| r^{2n} |\psi \rangle \approx \int_0^\infty \, dr \, r^{2n}
(u_\pi^2(r) + w_\pi^2(r)) + 2 \int_R^{R_\pi} \, dr \, r^{2n}
(u_\pi(r) \delta u (r) + w_\pi (r) \delta w(r)) + 
\delta \langle r^{2n} \rangle,
\end{equation}
where the first term is the object whose error, $\delta \langle r^{2n}
\rangle$, in the region $0 \leq r \leq R$, was assessed above.  The
second term is an additional error, however this error may be
bounded quite straightforwardly. Namely, since $V_{\rm TPE}$ is being
treated in perturbation theory the typical size of $\delta u$ will be
$O(P^2)$ relative to $u$. This suggests:
\begin{equation}
  \langle \psi| r^{2n} |\psi \rangle \approx \int_0^\infty \, dr \,
  r^{2n} (u_\pi^2(r) + w_\pi^2(r)) \, + \, \frac{2}{2n + 3} \, \,
  \gamma \left(R_\pi^{2n+1} - R^{2n + 1}\right) \,
  \frac{P^2}{\Lambda_\chi^2} + \delta \langle r^{2n} \rangle, 
\end{equation}
where we have made a somewhat simplistic assessment of the size of the
extra error by using the bounds derived above on the error if $u_\pi$ is
completely fallacious between $r=R_\pi$ and $r=R$, and then
multiplying that bound by the relative size $\delta u/u_\pi$.

In fact, the effects of two-pion exchange~ graphs will in general be
smaller than this, since their range is considerably less than that of
one-pion exchange~ mechanisms, and so additional exponential
suppressions will ameliorate the difference $u_\pi - u$ in the region
near $r=R_\pi$. In other words, two-pion exchange graphs not only
occur at a range which is shorter than that of one-pion exchange,
their effects are chirally suppressed relative to one-pion exchange at
the same distances by a factor $P^2/\Lambda_\chi^2$. This is an
argument based on chiral symmetry which lends support to the
conclusion of Ericson and Rosa-Clot that the effect of two-pion
exchange on the number extracted for $A_D/A_S$ is only five percent at
most, with about half of that coming from resonant two-pion exchange
in the rho-meson region~\cite{ERC83}.

There will also be corrections to one-pion exchange due to the $\pi
NN$ form factor. Any pionic corrections will be of one-pion range, but
will also be chirally suppressed, and so will not be dealt with here.
Indeed, Kaiser {\it et al.}~\cite{KW97} claim that the pionic corrections
give only mass and wave function renormalization corrections to
one-pion exchange up to $O(P^3)$ in the chiral expansion. At whatever
order such corrections do first contribute, their effects can be
allowed for in the same fashion in which we deal with relativistic
corrections to the one-pion exchange below.

Pionic corrections to the $\pi NN$ vertex are not normally included in
$NN$ potentials. One set of corrections which {\it are} usually
included are those due the nucleon's finite size. These are often
parametrized by a monopole form factor, leading to a momentum-space
one-pion exchange~ potential

\begin{equation}
  V({\bf q})=-\frac{4 \pi f_{\pi NN}^2}{m_\pi^2} \, \frac{{\bf \sigma}_1
    \cdot {\bf q} \, {\bf \sigma}_2 \cdot {\bf q}} {{\bf q}^2 +
    m_\pi^2} \, \left(\frac{\Lambda^2 - m_\pi^2}{\Lambda^2 + {\bf q}^2}
  \right)^2 \, (\vec{\tau}_1 \cdot \vec{\tau}_2).
\end{equation}
In fact, in coordinate space this potential gives central and tensor
potentials for the deuteron channel~\cite{ERC83}:

\begin{equation}
  V_C(r)=-m_\pi f_{\pi NN}^2 \left\{ \frac{e^{-m_\pi r} -
      e^{-\Lambda r}}{m_\pi r} - \frac{\Lambda^2 - m_\pi^2}{\Lambda
      m_\pi} e^{-\Lambda r} \right\}; 
\end{equation}
\begin{eqnarray} 
  V_T(r)= - m_\pi f_{\pi NN}^2 \left\{ \left[\frac{1}{m_\pi r} + 
    \frac{3}{(m_\pi r)^2} + \frac{3}{(m_\pi r)^3}\right] e^{- m_\pi r} 
- \left[\frac{3}{(m_\pi r)^3} + \frac{3 \Lambda}{m_\pi^3 r^2} +
    \frac{\Lambda^2}{m_\pi^3 r} \right] e^{-\Lambda r} \right. \nonumber\\
\left. - \frac{1}{2} \left(\frac{\Lambda^2}{m_\pi^2} - 1 \right)
\left(\frac{1}{m_\pi r} + \frac{\Lambda}{m_\pi} \right) e^{-\Lambda r}\right\}.
\end{eqnarray}
These potentials are exactly the same as Eqs.~(\ref{eq:VC}) and 
(\ref{eq:OPE5}) to a radius
$r \sim 1/\Lambda$.  Since $\Lambda$ is typically of order 1 GeV,
these corrections to one-pion exchange are almost irrelevant if OPEP
is only used for radii $r \geq 1.5$ fm. The effects of the $\pi NN$
form factor appear in the $\Lambda$-counting $NN$ potential as
higher-order corrections to (\ref{eq:VNN0})~\cite{Or96}.

\subsection{Relativistic corrections to the nucleon motion}

\label{sec-relativisticcorrections}

Relativistic corrections to both the potential $V$ and the 
free-nucleon propagation should also be considered. The latter
have no definite range. They are, however, suppressed by a factor:
\begin{equation}
\frac{\langle p^4\rangle}{8 M^3} \frac{1}{\gamma},
\end{equation}
and so we will make only corrections of this relative order to the
$\langle r^{2n} \rangle$'s. 

Relativistic corrections to the one-pion exchange~ potential can be
dealt with consistently in the chiral expansion. The leading effects
have the same chiral order as two-pion exchange. (See
Refs.~\cite{Or96,KW97} for details.)  Of course, in this case the
suppression relative to the leading one-pion exchange~ mechanism is due
to the smallness of the parameter $p/M$, rather than the (larger)
chiral parameter $m_\pi/\Lambda_\chi$. We see that corrections due to
relativistic pieces of one-pion exchange~ will make a contribution to
$\langle r^{2n} \rangle$:
\begin{equation}
  \delta \langle r^{2n} \rangle_{{\rm ROPE}} \approx \frac{2}{2n + 3}
  \, \gamma \left(R_\pi^{2n} - R^{2 n}\right) \, \frac{\langle p^2
    \rangle}{M^2}.
\label{momerr}\end{equation}

The key question is therefore, what is the size of the parameter
$\delta^2 \equiv \langle p^2 \rangle/M^2$?  In the pionless theory the
answer to this question is straightforward. The dominant contribution
comes from the piece of the wave function inside the short-distance
potential.  Hence the precise value of this operator is not
well predicted in the effective theory. Nevertheless, it is easy to see
that the quantity $\delta$ is indeed generically small.
Straightforward evaluation of $p^2$ with the wave function in the
pionless theory gives:
\begin{eqnarray}
  \langle p^2 \rangle &\approx& \alpha_0^2 \left( 1 - \frac{A_S^2}{2 \gamma}
  e^{-2 \gamma R_\pi}\right) \nonumber\\ 
  &\approx& 2 \, \alpha_0 \gamma \, (\alpha_0 R_\pi).
\end{eqnarray}
To avoid confusion we have denoted by $R_\pi$ the range to which the
short-distance potential extends in the {\it pionless} theory.  But
typically $\alpha_0 \sim 2/R_\pi$, so,

\begin{equation}
  \langle p^2 \rangle \approx 8 \frac{\gamma}{R_\pi},
\end{equation}
which, for a generic value, $R_\pi=3$ fm, is much less than $M^2$, thus
ensuring a small value of $\delta$ in the pionless theory.

In the theory with explicit pions we argue that the dominant
contribution again comes from the short-distance piece of the
potential. Thus in essence the only thing that changes for the above
argument when pions are added is that $R_\pi$ becomes $R$.  The
shortest $R$ we will consider, $R=1.5$ fm, which still only gives
$\langle p^2 \rangle^{1/2} \approx 220$ MeV.

Therefore although $\langle p^2 \rangle = M B$ does not hold, and this
quantum average is not well described in our effective field theory
approach to the deuteron, we may confidently say that relativistic
corrections to the deuteron electromagnetic current are suppressed.
The quantitative statement is that
\begin{equation}
  \frac{\langle p^{2n} \rangle}{M^{2n}} \, \, \approx \, \, 4 \, 
  \left(\frac{2}{MR}\right)^{2n - 1} \, \frac{\gamma}{M}.
\end{equation}
Note again that there is a crucial additional suppression by a factor
$\gamma/M$ due to the dominance of the tail region in the
shallowly-bound deuteron state.

\subsection{Estimates of $R$ from potential models}

\label{sec-estimates}

We now discuss typical values of the scale $R$. The Nijmegen
phase shift analyses of Refs.~\cite{St93,Re99} both suggest that
inside $r=1.4$ fm some dynamics other than two-pion exchange starts to
play a significant role~\footnote{As discussed previously, the
  requirement that the wave functions $u_\pi$ and $w_\pi$ result in a
  normalizable $\psi({\bf r})$ leads to a lower-bound on $R$ once a
  particular short-distance potential is chosen. Here that bound is
  1.3 fm. In the phase shift analysis of Ref.~\cite{St93} a similar
  bound, $R \geq 1.4$ fm, was found.}. Potential models based on
one-boson exchange interactions also have significant strength at $r
\approx 1.5$ fm which does not come from either one-pion exchange~ or
the two-pion exchange graphs in which the exchanged pions do not
interact. Effects such as ``correlated two-pion exchange'' are thought
to provide strength in this region and are parametrized in the
one-boson exchange interaction via (possibly broad) $\sigma$ and
$\rho$ mesons. The masses of these mesons are generally of order
500--800 MeV. Although their full effect will not be felt until
distances less than 0.5 fm are reached they will already have some
impact on the wave function at $r=1.5$ fm.  The low-momentum effects
of these heavier mesons will appear in the effective theory, but it is
not yet clear whether this will allow a good description of the
deuteron wave function at, say $r=1$ fm. Certainly, given the
low-order $NN$ interaction used here, for this work taking $R \geq
1.5$ fm is sensible.  Whether this distance can really be thought of
as being due to the hadronic scale $\Lambda_{\chi}$ is a delicate
matter which will engender some debate, since there is a factor of
three or four between $1/\Lambda_\chi$ and 1.5 fm. In particular,
Cohen and Hansen~\cite{CH98A,CH98B} have suggested that while an
expansion in $m_\pi/\Lambda_{\chi}$ is likely to be reasonable, an
expansion in $m_\pi R$ is questionable if $R$ is this large.

Finally, we can look at the plots in Fig.~\ref{fig-wfcomparison},
where we compare the results for the wave function $u_\pi(r)$ with
those from a typical phenomenological $NN$ potential, the Nijm93. We
see that for a range of $R$'s, from $R=1.5$ fm to $R=2.5$ fm, the wave
function $u$ and $w$ will be generically close to those from the much
more sophisticated Nijm93.  This agreement persists all the way down
to radii $r \approx 0.5$ fm.

For a number of reasons then, it seems that $R=1.5$ fm is a sensible
range to choose for the onset of short-distance physics in the $NN$
interaction which is beyond the scope of the effective theory.
However, we observe that, at least for the short-distance potential
used here, such a choice of $R$ leads to wave functions which track
closely with the ``true'' wave function to values of $r$ of order 0.5
fm. Thus, it may well be that the errors obtained when the wave
functions from such models are used will be significantly smaller than
the estimates derived above, since the range over which the error in
$\langle r^{2n} \rangle$ accrues is actually significantly less than
$R$.

\section{The electromagnetic current}

\label{sec-kernel}

Thus far our approach is not different to a primitive potential model
calculation of the deuteron wave function. Our deuteron wave function
has a tail which is constrained by the experimental data, a section at
one-pion range calculated from the one-pion exchange~ potential, and
some short-distance piece inside $r=R$, which here is produced by a
potential that is completely phenomenological. This is very similar to
the procedure adopted in the Nijmegen phase shift analyses of
Refs.~\cite{St93,Re99}. We are arguing that such a calculation is in
the spirit of effective field theory since we have separated the
effects that different scales in the problem, $\gamma$, $m_\pi$, and
$R$ have on the wave function. We now wish to demonstrate that we may
make truly systematic statements about observables. In this section we
will first explain how standard techniques of effective field theory
allow us to precisely quantify the validity of the nonrelativistic~
impulse approximation for the calculation of the deuteron
electromagnetic form factors. The goal is to show how the effect of
the known scales in the problem can be traced through so that the
impact of these scales on corrections to the calculations of measured
quantities can be known, even if the corrections themselves are not
calculated.

This allows us to calculate the electromagnetic current for the
deuteron, order-by-order in an expansion in $p$, $Q$ and $m_\pi$.  As
we shall see, it is a straightforward matter to make such an expansion
of the full single-nucleon currents. It is less straightforward to
obtain a systematic expansion for the two-body currents which
contribute to electron-deuteron scattering.  To do that we use
powerful effective field theory techniques to quantify the order in
the chiral expansion at which such corrections appear. These
techniques amount to using the constraints of the symmetries in the
problem (especially chiral symmetry) to determine which operators can
appear in the two-body piece of the current. This operator will then
have a definite dimensionality.  As we will explain below, the
assumption that the coefficient of this operator is natural gives a
direct estimate of the effect of the two-body mechanisms. The result
is an expansion for the electron-$NN$ scattering kernel based on naive
dimensional analysis, or $\Lambda$-counting, as developed by 
Weinberg~\cite{We90,We91,We92,Le99,Le97}.  

\subsection{Definitions}

\begin{figure}[h,t,b,p]
  \vspace{0.5cm} \epsfysize=5.5 cm \centerline{\epsffile{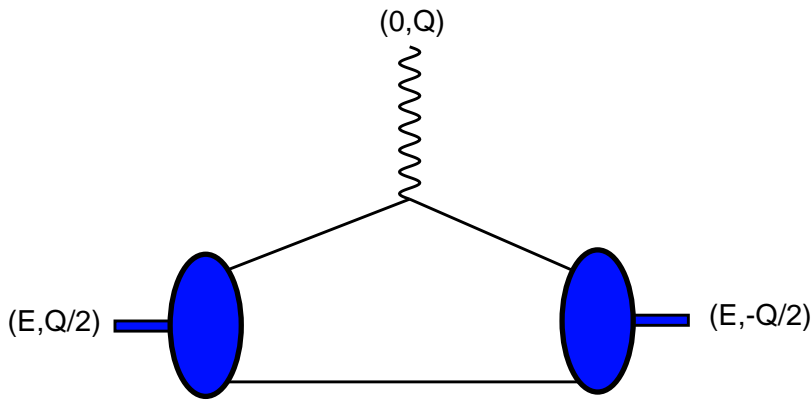}}
  \centerline{\parbox{11cm}{\caption{\label{fig-Breit} The kinematics
        for electron-deuteron scattering in the Breit frame. Only the
        impulse approximation diagram is shown.  }}}
\end{figure}

In this section we discuss the calculation of the electromagnetic 
form factors of the deuteron. Here we will work always in the Breit frame, 
where the kinematics are as displayed in Fig.~\ref{fig-Breit}. We
must calculate the matrix elements

\begin{equation}
\langle M' \, \, {\bf Q}/2| J_\mu | M \, \, {\bf -Q}/2\rangle
\end{equation} 
of the deuteron electromagnetic current to which a virtual photon of
three-momentum ${\bf Q}=Q \hat{z}$ couples, when the deuteron is in
states with specific magnetic quantum numbers $M$ and $M'$. In fact,
due to symmetry, only three of these matrix elements are
independent~\cite{deFW66}.  Thus, the electromagnetic structure of the
deuteron may be parametrized in terms of three form factors,
$F_C(Q^2)$, $F_M(Q^2)$, and $F_Q(Q^2)$, which are related to the Breit
frame matrix elements via:
\begin{eqnarray}
F_C(Q^2)&=&\frac{1}{3 e \eta_1}\left[
\langle 0 \, \, {\bf Q}/2| J_0 | 0 \, \, {\bf -Q}/2\rangle
+ \langle 1 \, \, {\bf Q}/2| J_0 | 1 \, \, {\bf -Q}/2\rangle
+ \langle -1 \, \, {\bf Q}/2| J_0 | -1 \, \, {\bf -Q}/2\rangle \right],\\
F_Q(Q^2)&=&\frac{2}{e Q^2 \eta_1}
\left[\langle 0 \, \, {\bf Q}/2| J_0 | 0 \, \, {\bf -Q}/2\rangle
- \langle 1 \, \, {\bf Q}/2| J_0 | 1 \, \, {\bf -Q}/2\rangle \right],\\
F_M(Q^2)&=&\frac{\sqrt{2} M_d}{e Q \eta_1} 
\langle + 1 \, \, {\bf Q}/2| J^+ | 0 \, \, {\bf -Q}/2\rangle;
\end{eqnarray}
where $\eta_1=\sqrt{1 + \frac{Q^2}{4 M_d^2}}$, and $J^+=J^1 + i J^2$.
  At $Q^2=0$ the form factors are normalized to:
\begin{eqnarray}
F_C(0)&=&1;\\
F_Q(0)&=&Q_d;\\
F_M(0)&=&\mu_d \frac{M_d}{M};
\end{eqnarray}
where $Q_d$ is the deuteron quadrupole moment, and $\mu_d$ is the magnetic
moment of the deuteron in units of nuclear magnetons. 

\begin{figure}[h,t,b,p]
  \vspace{0.5cm} \epsfysize=4.5 cm \centerline{\epsffile{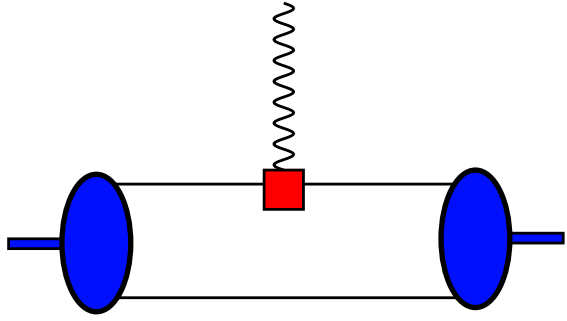}}
  \centerline{\parbox{11cm}{\caption{\label{fig-obc} A generic
        one-body current contribution to electron-deuteron scattering.
        Here the square indicates any $\gamma NN$ vertex, and the
        blobs are deuteron vertex functions.}}}
\end{figure}

\begin{figure}[h,t,b,p]
  \vspace{0.5cm} \epsfysize=4.5 cm \centerline{\epsffile{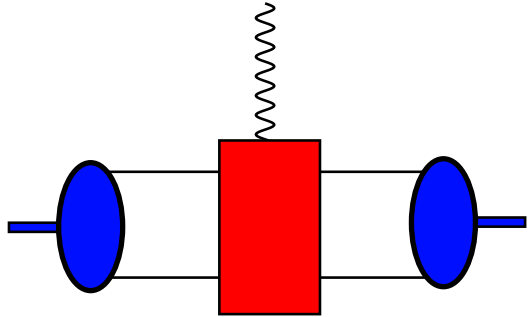}}
  \centerline{\parbox{11cm}{\caption{\label{fig-tbc} The two-body
        current contribution to electron-deuteron scattering. The
        rectangle is the two-body $NN \rightarrow NN \gamma$ kernel,
        which in this work is discussed in the framework of chiral
        perturbation theory. The blobs are deuteron vertex
        functions.}}}
\end{figure}

In general, the current to which the electron couples in
electron-deuteron scattering may be expressed as:
\begin{equation}
  \langle {\bf P}' \, \, {\bf p}'|J_\mu|{\bf P} \, \, {\bf p} \rangle
  =(2 \pi)^3 j_\mu^{(1)}({\bf p},{\bf Q}) \delta ({\bf p}' - {\bf p} - {\bf
    Q}/2) + j_\mu^{(2)}({\bf p},{\bf p}';{\bf Q}),
\end{equation}
where ${\bf p}$ and ${\bf p}'$ are the initial and final relative
momenta in the deuteron state, and ${\bf Q}={\bf P}' - {\bf P}$ is the
three-momentum of the photon exchanged between the electron and the
deuteron. The one and two-body currents discussed here are depicted in
Figs.~\ref{fig-obc} and \ref{fig-tbc}.

\subsection{One-body currents}

For on-shell nucleons the most general form that the single-nucleon
current, $j_\mu^{(1)}$, can take, consistent with Lorentz invariance,
time-reversal invariance, and current conservation, is:

\begin{equation}
  j_\mu^{(1)}({\bf p},{\bf Q})=e_N \bar{u}({\bf p} + 3 {\bf Q}/4)
  \left[F_1(Q^2) \gamma_\mu + F_2(Q^2) \frac{i}{2 M} \sigma_{\mu \nu}
    Q^\nu \right] u({\bf p} - {\bf Q}/4),
\label{eq:job}
\end{equation}
where $e_N$ is the nucleonic charge.  Note that we have labeled the
momenta as they would occur were we to calculate electron-deuteron
scattering in the Breit frame and then look at the single-nucleon
current as a function of the relative momentum in the deuteron, ${\bf
  p}$.  For the deuteron case, $F_1$ and $F_2$ are isoscalar
combinations of the proton and neutron Dirac and Pauli form factors:
\begin{eqnarray}
F_1(Q^2)&=&F_1^{(p)}(Q^2) + F_1^{(n)}(Q^2);\\
F_2(Q^2)&=&F_2^{(p)}(Q^2) + F_2^{(n)}(Q^2).
\end{eqnarray}

Now we expand $j_\mu^{(1)}$ in powers of the momenta ${\bf p}$ and ${\bf Q}$.
This involves expanding out both the Dirac structures which
arise from Eq.~(\ref{eq:job}), and the single-nucleon form
factors $F_1(Q^2)$ and $F_2(Q^2)$. The first expansion is governed
by ratios of momenta to the nucleon mass, which we denote generically
by the mnemonic:

\begin{equation}
\delta^2 \, \, \sim \, \, \frac{p^2}{M^2} \, , \, \frac{p \cdot Q}{M^2} \, ,
\, \frac{Q^2}{M^2}.
\end{equation}
Corrections at order $\delta^2$ include the usual spin-orbit and
Dirac-Foldy relativistic corrections to the charge
operator~\cite{Fr73}. Meanwhile, the second expansion will be
governed by the parameter $Q^2/\Lambda_{\chi}^2$. It is not surprising that 
when working to one order beyond the lowest nontrivial
contributions, the current is simply due to two objects of charges
one and zero, and to anomalous magnetic moments $\mu_p$ and $\mu_n$.
Corrections due to the neutron and proton radii appear at the next
order:
\begin{eqnarray}
  j_0^{(1)}({\bf p},{\bf Q})&=& e \left(1 + O(\delta^2) +
    O\left(\frac{Q^2}{\Lambda_\chi^2}\right)\right),\\ j_+^{(1)}({\bf
    p},{\bf Q})&=&e \left( \frac{p_+}{M} + \frac{Q}{2M} (\mu_p + \mu_n
    + 1) \sigma_+ + O(\delta^3) + O\left(\delta
      \frac{Q^2}{\Lambda_{\chi}^2}\right) \right). \label{eq:jplus}
\end{eqnarray}

It should be noted that due to reparametrization invariance any
nonrelativistic~ chiral Lagrangian must reproduce exactly the
corrections in $\delta$ found by expanding out Eq.~(\ref{eq:job}). 
In such an approach the corrections in $Q^2/\Lambda_{\chi}^2$ 
will appear as higher-derivative operators of the type $N^\dagger
\nabla^2 A_\mu N$.  These operators have unknown coefficients which
must be fit from the experimental data.  The reader who wishes to see
an analysis of the single-nucleon current carried out in this
completely proper fashion should refer to Ref.~\cite{Be98}. Note that
examination of the proton form factor data suggests that the expansion
of the single-nucleon form factors in powers of $Q$ will break down
completely at a scale $Q$ of order 800 MeV, and will begin to fail
well before that.

We will take the expressions for $j_0$ and $j_+$ only at
the leading non-vanishing order in each case. The corrections to these
expressions are then suppressed by at least two powers of $\delta$.
Above we showed that $\langle p^2 \rangle/M^2$ was a generically small
quantity for the wave functions discussed here, and therefore we can be
confident that these NNLO corrections  are indeed higher-order effects
provided that $Q < 1$ GeV.

\subsection{Two-body currents}

We are now faced with the task of quantifying the effect of two-body
currents on the observables under consideration. It is here that the
chiral counting approach advocated by Weinberg~\cite{We90,We91,We92},
allows us to make real headway. For a review of this approach
see~\cite{vK99}.  Here we just restate and employ the results which we
need to proceed.

\begin{figure}[h,t,b,p]
  \vspace{0.5cm} \epsfysize=3.75 cm \centerline{\epsffile{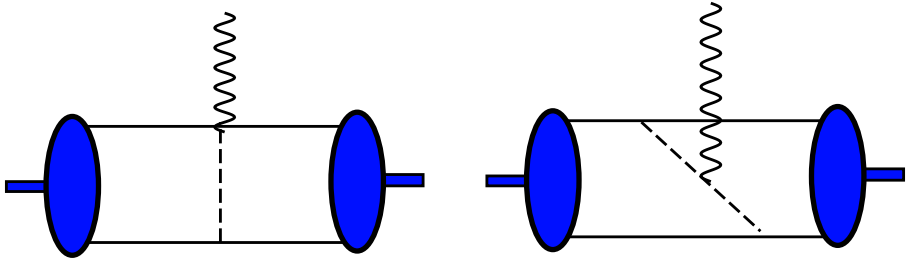}}
  \centerline{\parbox{11cm}{\caption{\label{fig-tbcpions} The contribution
        to electron-deuteron scattering from the leading-order pionic
        current.  Here all vertices are from the leading-order chiral
        perturbation theory Lagrangian. The graphs with nucleons one
        and two interchanged also contribute.}}}
\end{figure}

The contribution of two two-body pionic mechanisms to the deuteronic
current is depicted in Fig.~\ref{fig-tbcpions}.  It is straightforward
to count the powers of momentum or pion mass in such graphs. If we
generically label internal and external momenta and the pion mass as
$P$, i.e. write $p,Q,m_\pi \sim P$, we then have the following rules
for the scaling of a given contribution to the electron-deuteron
scattering kernel:

\begin{itemize}
\item Each two-nucleon propagator scales like $1/P^2$;

\item Each nonrelativistic~ loop integral contributes a factor of $P^3$;

\item Each pion propagator scales like $1/P^2$;

\item Each vertex contributes {\it at least} one power of $P$0.
\end{itemize}

The key point here is the last one. It arises because the
spontaneously broken chiral symmetry of QCD implies that pions couple
derivatively to the nucleon. Thus the current contributions in
Fig.~\ref{fig-tbcpions} scale {\it at least} as $P^2$.~\footnote{Here
  we are counting the charge $e$ as of order $P$ in order to
  simplify matters.}.  Therefore we expect that such meson-exchange
currents will be of order
\begin{equation}
  \frac{m_\pi^2}{\Lambda_\chi^2} \, \, , \, \, \frac{\langle p^2
    \rangle}{\Lambda_\chi^2} \, \, , \, \, \frac{Q^2}{\Lambda_\chi^2} \, \, ,
  \, \, \frac{\langle p \rangle \, Q}{\Lambda_\chi^2}.
\end{equation}

\noindent This is to be compared to the contribution in Fig.~\ref{fig-obc} when
the vertex is taken to be the lowest-order $\gamma NN$ interaction.
In that case Fig.~\ref{fig-obc} has one less loop, one less propagator
and one less vertex than Fig.~\ref{fig-tbcpions}, and so naively
scales as $P^0$. In fact, Eq.~(\ref{eq:jplus}) suggests that magnetic
coupling is a sub-leading effect in this expansion; and therefore if
$j_+$ is used at the $\gamma NN$ vertex in Fig.~\ref{fig-obc} then the
overall graph is order $P$.

This might lead one to think that meson-exchange currents are an NLO
effect in the deuteron form factor $F_M$, but this is not the case. In
fact, although the leading effect of meson-exchange currents is
nominally $O(P^2)$, the meson-exchange current which occurs at that
order is isovector in character and thus vanishes in the deuteron. 
It is well known that this 
this $O(P^2)$ isovector current is the leading MEC correction to
processes such as electrodisintegration of the deuteron
(see, for instance, Ref.~\cite{Ri84}). In practice this results in the
leading effect from meson-exchange currents coming at $O(P^3)$, or
NNNLO in the deuteron electric form factor and NNLO in the magnetic
form factor. In such an $O(P^3)$ exchange current the $\gamma \pi$
contact interaction with the nucleon has a coefficient that is related
to known constants by the requirements of gauge invariance and
reparametrization invariance. We note that this correction to the
charge operator is phenomenologically important in potential-model
calculations of $F_C$~\cite{CS98}.

At some order above $O(P^2)$ operators enter whose coefficient is not
constrained by symmetries, but instead is sensitive to short-distance
physics. One such mechanism will be the analogue of the $\rho \pi
\gamma$ MEC employed in some hadronic model calculations of
electron-deuteron scattering. 

Finally, in considering meson-exchange current corrections to $F_Q$
it is important to note that the natural size of $F_Q$ at $Q=0$ is 
much smaller than that of $F_C$ and $F_M$. A natural expectation for 
$\left. F_Q\right|_{Q=0} \equiv Q_d$ is:

\begin{equation}
Q_d \sim \frac{\eta}{\gamma^2}.
\end{equation}
Here $\eta$---the asymptotic D-to-S normalization---enters because an
S-wave deuteron cannot emit a quadrupole photon via any one-body
current mechanism. Thus the power-counting for the contribution of
one-body currents to $F_Q$ is exactly the same as for similar
contributions to $F_C$. However, two-body currents can connect the S-wave
pieces of the deuteron initial- and final-state wave functions one to
another and still result in the emission of a quadrupole
photon~\cite{Ch99}. The first such effect does not occur until
$O(P^5)$ in $\Lambda$-counting. On the other hand, crucially it contains no
powers of $\eta$. Hence, numerically it can be important at much lower
order than naively expected. A quick check shows that the contribution
of this operator to $Q_d$ will be of order

\begin{equation}
  \Delta Q_d \, \sim \, \frac{\langle p^3 \rangle}{M \Lambda^2} \, \, 
  \frac{1}{\Lambda^2}.
\end{equation}
Although this is formally of order $P^5$, numerically it is as important
as NNLO corrections to $Q_d$.

In summary, our expansion in powers of $P$ for the one- and two-body
currents which contribute to electron-deuteron scattering leads us to
conclude that the usual nonrelativistic~ one-body currents for
pointlike nucleons will give the correct result up to corrections of
NNLO in both $F_C$ and $F_M$.  At that order relativistic effects will
correct the one-body current and effects due to electromagnetic form
factors of the nucleon will arise. Meson-exchange currents may also
contribute to $F_M$ at that order. The first meson-exchange
contribution to $F_C$ occurs at NNNLO.

\section{Results}

From Sec.~\ref{sec-wavefunction} we have a variety of deuteron
wave functions with different short-range behavior. From
Sec.~\ref{sec-kernel} we have an understanding of the deuteron
current in a systematic $\Lambda$-expansion. Therefore in this section
we proceed to calculate the electromagnetic form factors of the
deuteron. This is done for various values of $R$ ranging from 1.5 fm
to 3.5 fm in the theory with pions, and for the minimum $R$ ($R=2.0$ fm)
in the pionless theory. These results are compared with results from a
sophisticated potential model, the Nijm93 potential. In every case we
employ the nonrelativistic~ impulse approximation with point nucleons,
which, as explained above, represents a calculation to NLO in the
$\Lambda$-counting electron-deuteron scattering kernel.  However,
before calculating these form factors we spend some time discussing
our results for static properties of the deuteron.

\subsection{Static moments}

The values of the form factors $F_M$ and $F_Q$ at $Q^2=0$ are related
to the magnetic and quadrupole moments of the deuteron. It is well
known that to this order in $\Lambda$-counting these may be obtained
directly as simple integrals of the deuteron radial wave functions $u$
and $w$:
\begin{eqnarray}
  \mu_d&=&\mu_p + \mu_n - \frac{3}{2}(\mu_p + \mu_n - \frac{1}{2}) P_D;\\ 
  Q_d&=&\frac{1}{\sqrt{50}}\int_0^\infty dr \, r^2 \, u(r) w(r) -
  \frac{1}{20} \int_0^\infty dr \, r^2 \, w^2(r)
\end{eqnarray}
with the D-state probability $P_D$ given by:
\begin{equation}
P_D=\int_0^\infty dr \, w^2(r).
\end{equation}
Employing the wave functions derived above we obtain the results shown
in Table~\ref{table-static}. 

\begin{table}[h,t,b,p]
\label{table-static}
\begin{center}
\begin{tabular}{|c|c|c|c|}
\hline
Theory       & Value of R (fm)  &  $\mu_d$ ($\mu_N$) & $Q_d$ (${\rm fm}^2$)  \\ \hline \hline
Nijm93       &      N/A    &  0.847     &   0.271 \\ \hline
Pionless     &      2.0   &  0.804     &   0.322 \\ \hline
Pionful      &      1.5    &  0.851     &   0.269 \\ \hline
Pionful      &      2.0   &  0.858     &   0.265 \\ \hline
Pionful      &      2.5    &  0.863     &   0.258 \\ \hline
Pionful      &      3.5    &  0.871     &   0.238 \\ \hline
\end{tabular}
\end{center}
\caption{Static moments of the deuteron for various values of $R$ 
in the pionless and pionful theories.}
\end{table}

In Table~\ref{table-static} we clearly see that the deuteron D-state
is much too large in the only deuteron calculated in the pionless
theory, while it is too small in all the calculations where pions were
included explicitly in the potential---albeit only marginally so for
$R=1.5$ fm. Examination of the deuteron wave functions shows that when
integrating in from infinite $r$ the tensor piece of one-pion exchange
causes the deuteron D-state wave function to turn over. Without this
effect in the calculation the deuteron D-state becomes too large.
Once this effect of one-pion exchange is included making $R$ large
causes the D-state wave function to turn over too early, thereby
leading to a reduction of the D-state probability. Finally, we note
that if $R$ is made very large the pionful calculation begins to
resemble the pionless calculation, since the one-pion exchange
potential is too weak in the region $r > R$ to have much effect on the
wave function. These observations echo the work of Ericson and
Rosa-Clot, who attempted to use methods like the ones we are employing
here to put constraints on $\eta$ from the experimental value of
$Q_d$~\cite{Er84,ERC85}.

This significant variation in $Q_d$ shows that there is some
sensitivity to the short-distance physics in this observable. Indeed,
it should be noted that the experimental value for this quantity is
actually $Q_d=0.2859(3) {\rm fm}^2$~\cite{BC79}, rather than the value
$Q_d=0.271 \, {\rm fm}^2$ obtained using the nonrelativistic~ impulse
approximation and the Nijm93 wave function. Apparently there are
significant contributions to this quantity from physics beyond the
nonrelativistic~ impulse approximation. Corrections to this
approximation appear at higher orders in the $\Lambda$-expansion of
the kernel than we have considered here. One correction is a two-body
current resulting in the emission of a quadrupole photon, the
Lagrangian for which was given in Ref.~\cite{Ch99}.  This mechanism is
of much higher order than the others we have considered in this work,
however it still gives approximately as large a contribution to $Q_d$
as terms one order beyond those we have considered here. The
importance of this higher-order correction to $Q_d$ is entirely due to
the smallness of the leading-order contribution to this observable. In
contrast, we shall see below that $F_C$ and $F_M$, which are not
unusually small quantities, really are insensitive to details of the
short-distance physics of the $NN$ potential at radii $r < 2.5$ fm,
provided that $Q$ is less than about 700 MeV.  But before
launching that investigation, we discuss the static moments $\langle
r^2 \rangle$ and $\langle r^4 \rangle$ which are dominated by the
long-distance piece of the deuteron wave function.

A conservative prediction for the error in these quantities as a
function of $R$ was made in the previous section. In
Table~\ref{table-rn} we display results for these quantities for
different values of $R$ in both the pionless and pionful calculations.
These moments are to be compared to those obtained with the Nijm93
deuteron wave function.  The behavior of the moments is clearest in
the pionless theory.  As expected, the error in $\langle r^4 \rangle$
is smaller, and grows more rapidly with $R$ than in $\langle r^2
\rangle$. For all but the extreme case of the pionless theory with $R
\geq 5$ fm the error is on the order of a few percent~for $\langle
r^2 \rangle$, and is less than one percent~for $\langle r^4 \rangle$.
Thus there is good agreement between these quantities as calculated
with our EFT-motivated wave functions and with the more sophisticated
Nijm93 wave function. The reason for this is clear: these are
tail-dominated observables, and we have made sure that the tail of all
of our wave functions agrees almost exactly with that of the
potential-model wave function.

\begin{table}[h,t,b,p]
\label{table-rn}
\begin{center}
\begin{tabular}{|c|c|c|c|}
\hline
Theory       & Value of R (fm)  &  $\langle r_m^2 \rangle$ (${\rm fm}^2$) 
& $\langle r_m^4 \rangle$ (${\rm fm}^4$)\\ \hline \hline
Nijm93       &      N/A    &  3.865    & 55.35  \\ \hline
Pionful      &      1.5    &  3.850    & 54.85  \\ \hline
Pionful      &      2.0    &  3.848    & 54.85  \\ \hline
Pionful      &      2.5    &  3.859    & 54.87  \\ \hline
Pionful      &      3.5    &  3.956    & 55.11  \\ \hline
Pionless     &      2.0    &  3.940    & 55.08  \\ \hline
Pionless     &      2.5    &  3.919    & 55.06  \\ \hline
Pionless     &      3.5    &  3.980    & 55.23  \\ \hline
Pionless     &      5.0    &  4.384    & 57.15  \\ \hline
\end{tabular}
\end{center}
\caption{Values of the matter radii 
  expectation values $\langle r_m^2 \rangle$ and $\langle r_m^4
  \rangle$ in the pionless and pionful theories for various values of
  $R$. Note that if $r$ is the relative coordinate between
  the two nucleons then $r_m=r/2$.}
\end{table}

\subsection{Electromagnetic deuteron form factors}

These results give us confidence that in the region where a moment
expansion of the deuteron electromagnetic form factors is valid these
form factors will be well reproduced by EFT calculations.
Furthermore, it means that within the domain of validity of that
expansion the error in the form factor can be estimated within our
calculational scheme. However, both of these results are of limited
use, since an expansion of the electromagnetic form factors of the
deuteron in moments of the charge distribution is only valid for $Q^2
\leq 4 / \langle r^2 \rangle$; i.e., $Q$ less than of order $\gamma$.

Of course, since the  wave functions found by integrating in agree
almost exactly with the Nijm93 wave function for distances $r$ greater
than $R$ we expect the electromagnetic form factors
of the deuteron obtained with these wave functions to be in
good agreement with those found with the Nijm93 wave function
up to $Q$ of order $2/R$. This is a much higher scale 
than $\gamma$; thus the breakdown of the moment expansion
does not mean that an effective field theory description of the 
deuteron is invalid.

\begin{figure}[h,t,b,p]
  \vspace{0.5cm} \epsfysize=9cm \centerline{\epsffile{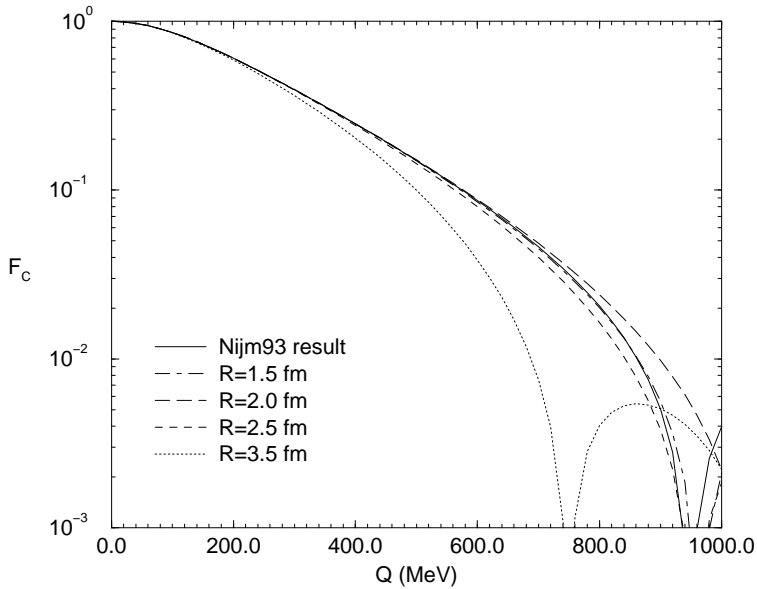}}
  \centerline{\parbox{11cm}{\caption{\label{fig-FCpionful} The charge
        form factor of the deuteron for several different wave
        functions, all of which have the same tail, as well as the same
        one-pion exchange part. The solid line is the result for the
        Nijm93 potential, while our wave functions with different R's
        are represented by R=1.5 fm (dot-dashed), R=2.0 fm (long
        dashed), R=2.5 fm (short dashed), and R=3.5 fm (dotted).}}}
\end{figure}

Turning first to the charge form factor of the deuteron we display
results in the pionful theory in Fig.~\ref{fig-FCpionful}. These are
compared with results from the Nijm93 wave function. The agreement is
good out to $Q$ of around 700 MeV, for all but the case $R=3.5$ fm. As
discussed above, that wave function is in fact very close to the one
calculated without explicit pions at the same $R$, so it cannot really
be considered a calculation ``with pions''. Thus, in practice we do a
little better than $Q \sim 2/R$ in the calculations with explicit
one-pion exchange. The scale of the breakdown is, for sensible $R$, more
like $Q \sim m_\rho$.

\begin{figure}[h,t,b,p]
  \vspace{0.5cm} 
  \epsfysize=9cm
  \centerline{\epsffile{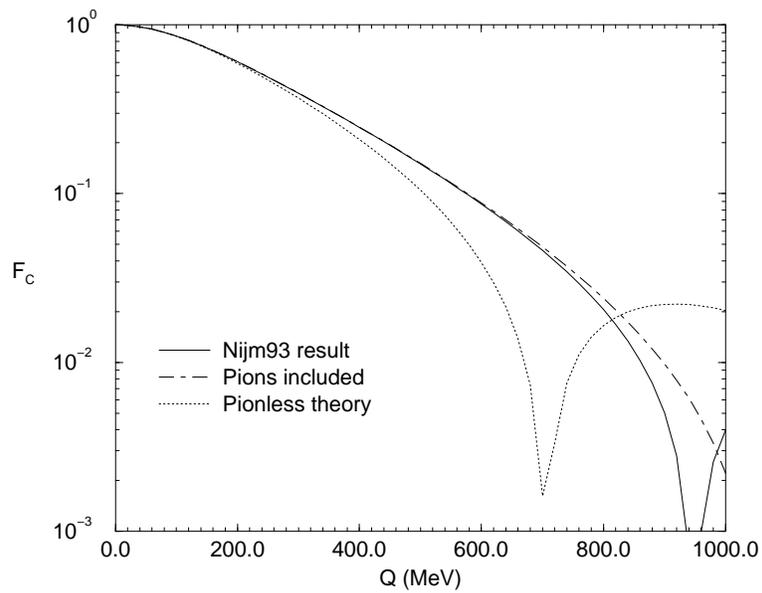}}
  \centerline{\parbox{11cm}{\caption{\label{fig-FCpionless} The charge
        form factor of the deuteron for a short-distance potential of
        $R=2.0$ fm, both with and without one-pion exchange included.
        The solid line is the Nijm93 result, the dotted line is the
        pionless calculation, and the dot-dashed line is a calculation
        including one-pion exchange.}}}
\end{figure}

What is the scale of the breakdown of the pionless theory? To establish
this, in Fig.~\ref{fig-FCpionless} we present results for $F_C$ from
the pionless and pionful theories with $R=2.0$ fm (the minimum
possible $R$ in the pionless case), and also for the Nijm93 potential.
From this plot it is clear that the pionless calculation breaks down a
little above $Q=m_\pi$, as we would expect.  When included as we have
done here, one-pion exchange makes a dramatic difference to the range
over which the description of the deuteron charge form factor is
successful.

\begin{figure}[h,t,b,p]
  \vspace{0.5cm} \epsfysize=9cm \centerline{\epsffile{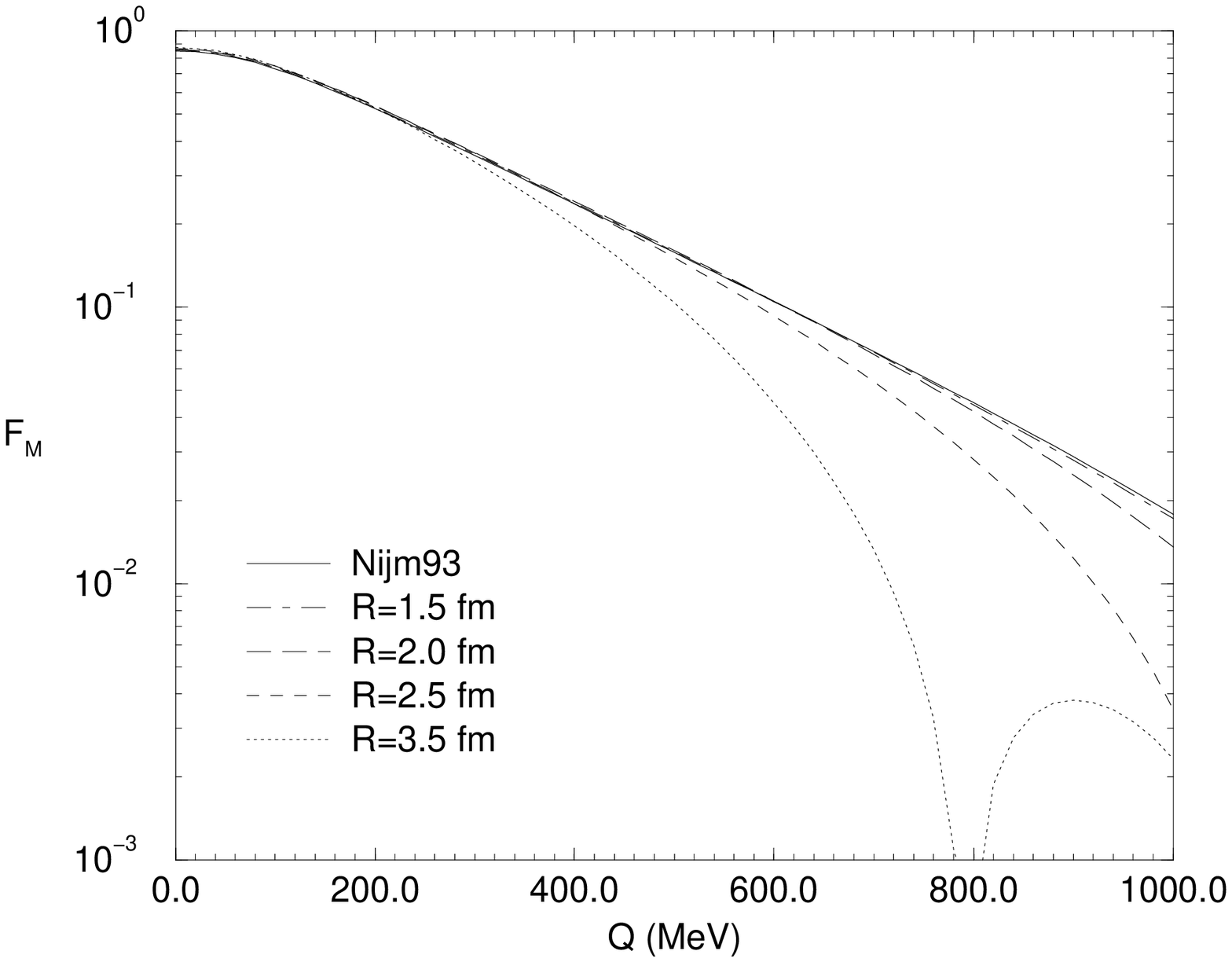}}
  \centerline{\parbox{11cm}{\caption{\label{fig-FMpionful} The magnetic
        form factor of the deuteron for several different wave
        functions, all of which have the same tail, and also the same
        one-pion exchange part. Legend as in Fig.~\ref{fig-FCpionful}.}}}
\end{figure}

\begin{figure}[h,t,b,p]
  \vspace{0.5cm} 
  \epsfysize=9cm
  \centerline{\epsffile{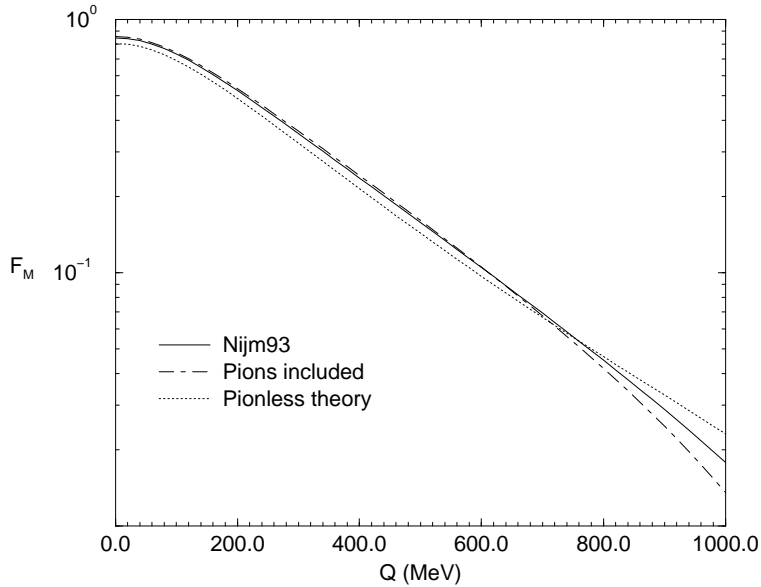}}
  \centerline{\parbox{11cm}{\caption{\label{fig-FMpionless} The magnetic
        form factor of the deuteron for a short-distance potential of
        $R=2.0$ fm, both with and without one-pion exchange included.
        Legend as in Fig.~\ref{fig-FCpionless}.}}}
\end{figure}

Results are presented for the deuteron magnetic form factor $F_M$ in
Figs.~\ref{fig-FMpionful} and \ref{fig-FMpionless}. These plots lead
to conclusions similar to those inferred from the plots of the charge
form factor. This is not particularly surprising, given that it was
shown in Ref.~\cite{Ka98C} that to leading order in the Q-expansion
the magnetic form factor is directly proportional to the charge form
factor. Of course, such a result is basically a consequence of the
dominance of the S-wave pieces of the wave function in determining the
magnetic structure of the deuteron at low momenta.

\begin{figure}[h,t,b,p]
  \vspace{0.5cm} \epsfysize=9cm \centerline{\epsffile{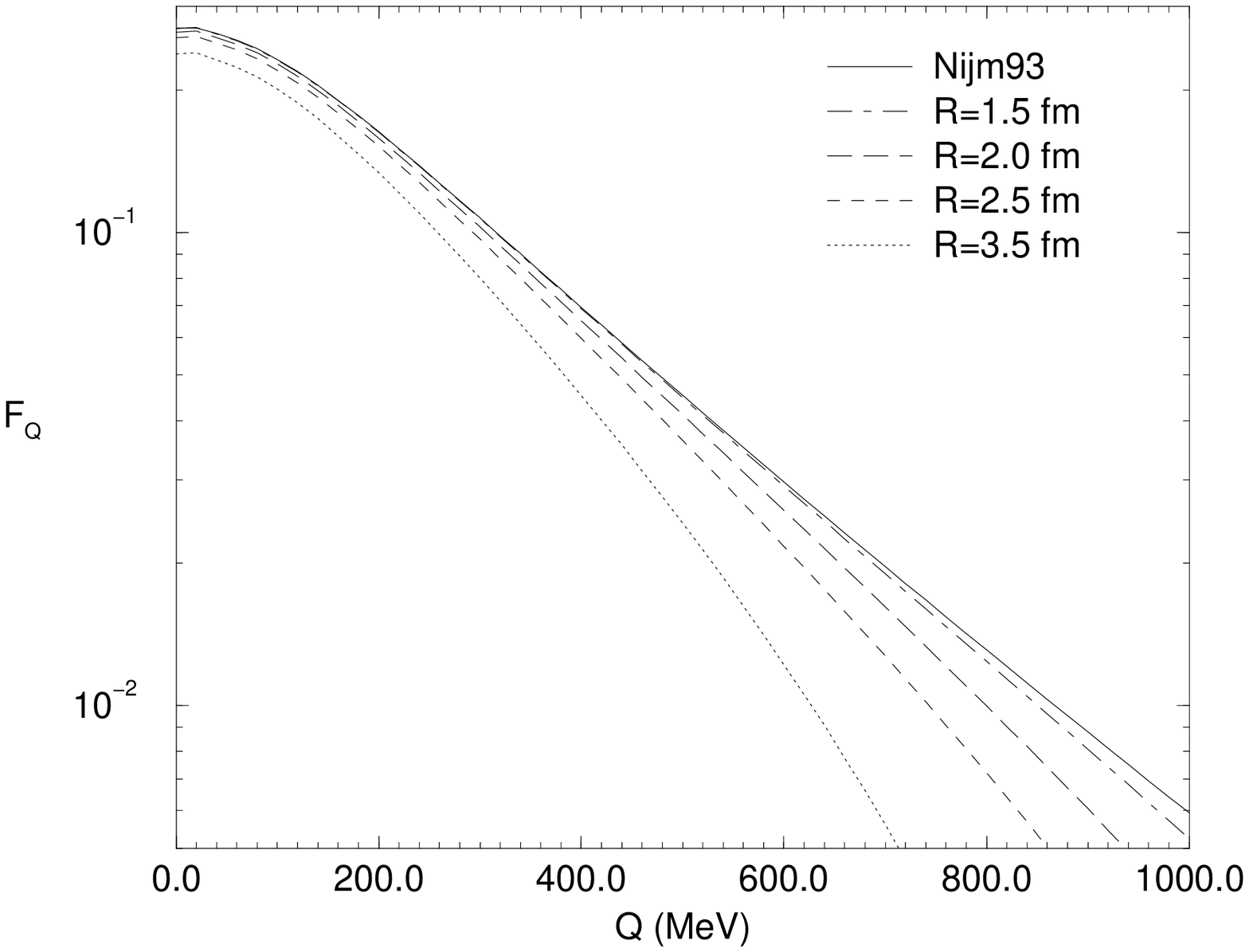}}
  \centerline{\parbox{11cm}{\caption{\label{fig-FQpionful} The quadrupole
        form factor of the deuteron for several different wave
        functions, all of which have the same tail, and also the same
        one-pion exchange part. Legend as in Fig.~\ref{fig-FCpionful}.}}}
\end{figure}

The situation is somewhat different for the quadrupole form factor
$F_Q$. As displayed in Fig.~\ref{fig-FQpionful} there is a problem
because the calculations with $R=2.5$ fm and $R=3.5$ fm do not
reproduce the quadrupole moment of the Nijm93 calculation very well.
On the other hand, the shape of all curves except the $R=3.5$ fm one is
basically correct out to about $Q=500$ MeV. The reason for this
discrepancy is clear if one examines the D-state wave functions for
$R=2.5$ fm. This wave function is a very poor approximation to the
Nijm93 wave function for $R < 2.5$ fm, although it agrees with it
almost exactly for $R \geq 2.5$ fm.  Thus, we expect that the
quadrupole form factor will be poorly described by such a wave
function once $Q$ is of the order of 300--400 MeV. 

\begin{figure}[h,t,b,p]
  \vspace{0.5cm} \epsfysize=9cm \centerline{\epsffile{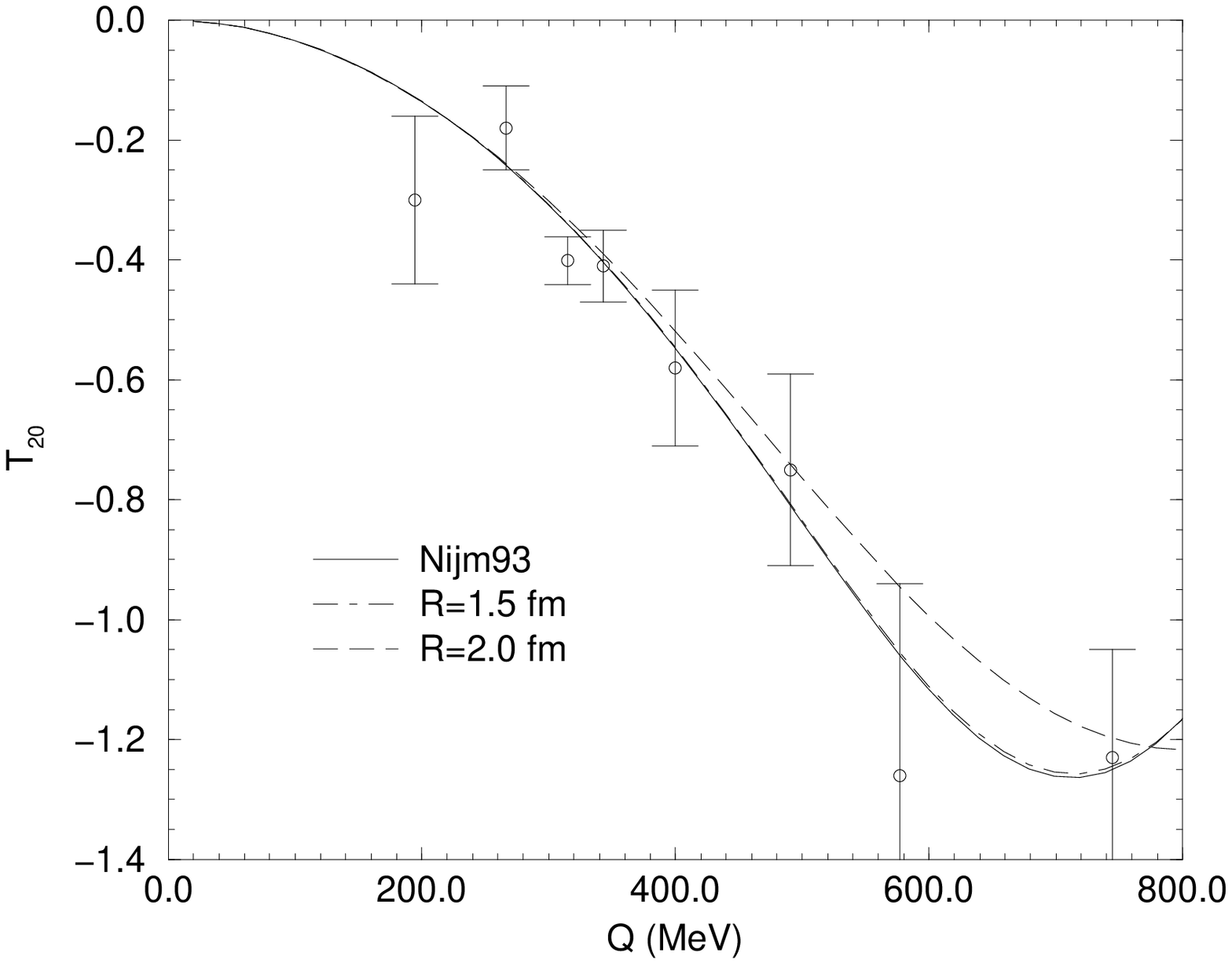}}
  \centerline{\parbox{11cm}{\caption{\label{fig-T20} The tensor
        polarization observable $T_{20}$. The solid line is the result
        using the Nijm93 wave function, while the dot-dashed and
        long-dashed lines are calculations with R=1.5 fm and R=2 fm
        respectively.}}}
\end{figure}

However, it should be noted that one can get a larger range of $Q$
over which different calculations agree if one restricts $R \leq 2$ fm.
In this case the low-momentum experimental data on the
electron-deuteron tensor-polarization observable $T_{20}$ is well
reproduced by wave functions with a range of different $R$'s. From
Fig.~\ref{fig-T20} it would seem that our lowest-order EFT
calculation can, provided $R$ is kept in an admittedly fairly small
range, describe the data of Refs.~\cite{Sc84}--\cite{Fe96}
fairly well out to about $Q \approx 700$ MeV.

Note that we have not compared our results for $A$ and $B$ with
experimental data. A brief look at our calculation suffices to show
that the scale of breakdown in predicting $A$ and $B$ will have little
to do with the nuclear physics of the problem. Instead, it will be set
by the assumption of pointlike nucleons.  Thus, calculations for $A$
and $B$ already do poorly with respect to the experimental data at $Q
\approx 400$ MeV. This failure is not due to sensitivity to
short-distance nuclear dynamics. Rather, it is due to the dominance of
nucleon-structure corrections in the higher-order contributions to the
$\Lambda$-counting kernel for this process.  In contrast, in $T_{20}$
the form factors of the nucleons essentially cancel out, and so it {\it is}
worthwhile to confront our NLO predictions with experimental data
there~\footnote{We thank Rocco Schiavilla for pointing this out to us,
  and for suggesting we calculate $T_{20}$ in this approach.}.

\begin{figure}[h,t,b,p]
  \vspace{0.5cm} 
  \epsfysize=9cm
  \centerline{\epsffile{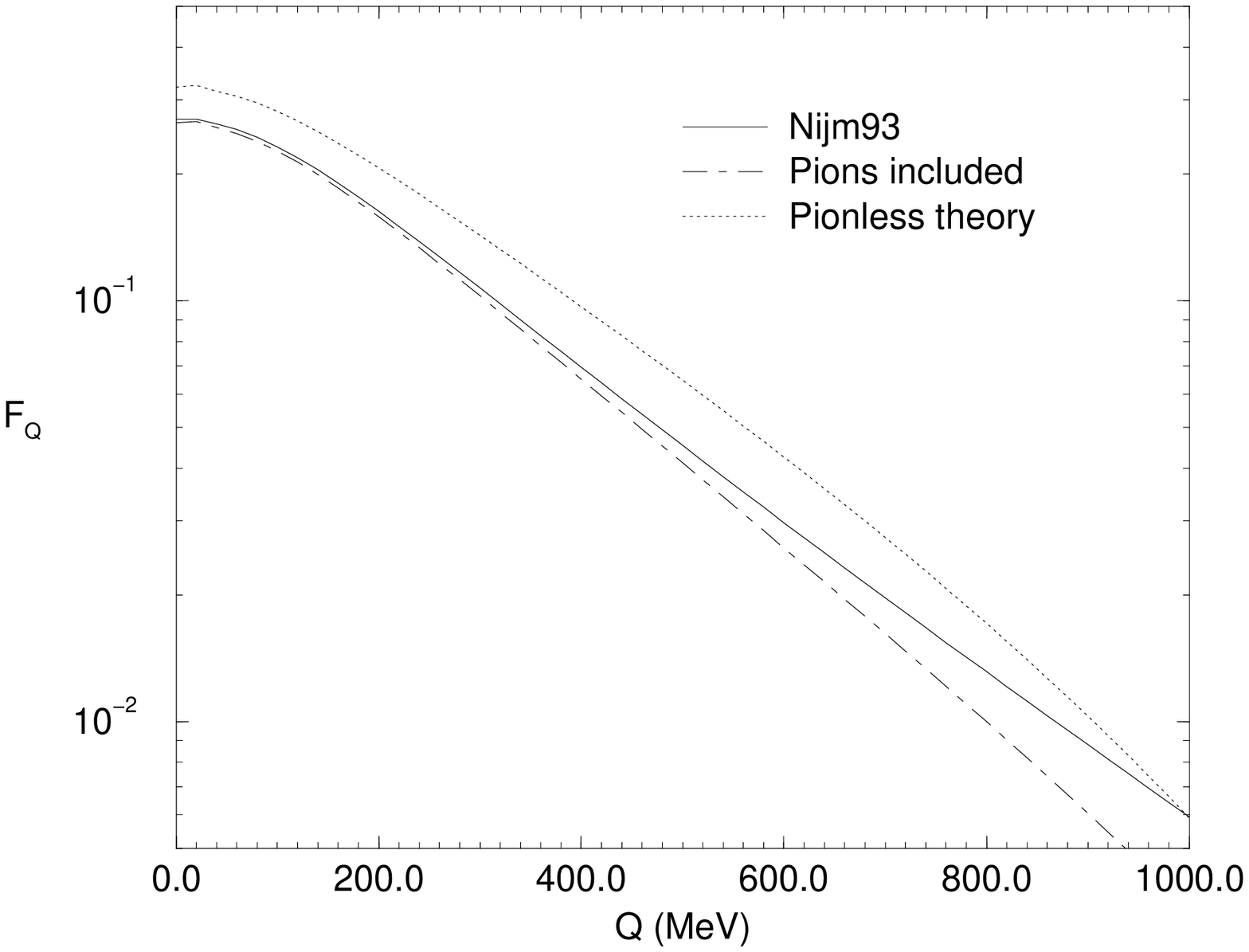}}
  \centerline{\parbox{11cm}{\caption{\label{fig-FQpionless} The quadrupole
        form factor of the deuteron for a short-distance potential of
        $R=2.0$ fm, both with and without one-pion exchange included.
        Legend as in Fig.~\ref{fig-FCpionless}.}}}
\end{figure}

Finally, Fig.~\ref{fig-FQpionless} shows results in the pionless
theory for $F_Q$. Since no tensor mixing is included in the pionless
calculation the only way that the D-state enters is through the
asymptotic D-to-S ratio. This physics is apparently not enough to
reproduce $Q_d$, although the low-momentum shape of $F_Q$ is
essentially correct. Similar results to this have been seen in NNLO
calculations of $F_Q$ using the version of Q-counting which applies in
the effective field theory without pions~\cite{Ch99}.

\section{Summary and discussion}

\label{sec-conclusion}

In this paper we have performed an effective field theory analysis of
electron-deuteron scattering. In such an approach the Lagrangian of
heavy-baryon chiral perturbation theory, which contains nucleons,
pions, and photons as explicit fields, is employed to derive a
diagrammatic expansion for any process. The naive engineering
dimension of these diagrams can then be computed and this allows a
particle-number-conserving Hamiltonian for the $NN$ system to be
written down as an expansion in operators of increasing
dimensionality. Then, if an operator has dimensionality $D$ relative
to the leading contribution to some observable its physical effects
are suppressed by $(P/\Lambda)^D$, where $P$ is either the typical
momentum in the problem, or $m_\pi$, and $\Lambda$ is either the mass
of the nucleon, or $\Lambda_\chi$, the chiral symmetry breaking scale.
This leads to an expansion for the $NN$ potential in which the leading
interaction is the sum of a one-pion exchange~ potential and a contact
interaction~\cite{We90,We91,Or96,Le99}.  Corrections to this are
suppressed by $P^2/\Lambda^2$.

In this work we implemented the ideas of this ``$\Lambda$-counting''
by demanding that the asymptotic deuteron wave function correspond to
the correct deuteron binding energy and the experimental values of
$A_S$ and $A_D$ (the asymptotic S- and D-state normalizations). This
wave function was then integrated in from $r=\infty$ to finite $r$,
using a Schr\"odinger equation which included the one-pion exchange~
potential.  At an arbitrary radius $R$ we then imposed a
short-distance regulator, whose role is to give the deuteron wave
function the appropriate $r \rightarrow 0$ limit. For $r>R$
corrections to the wave function thus obtained occur from higher-order
effects in the $\Lambda$-counting expansion of the $NN$ potential, and
so are suppressed by $P^2/\Lambda^2$. Effective field theory then
leads us to hope that physical observables will not be sensitive to
the artificial short-distance potential we have imposed. Indeed, the
statement that observables should not depend on $R$ is really the
renormalization group for this formulation of effective field theory,
with $R$ playing the role of an inverse cutoff. The degree to which we
have renormalizaton-group invariance, i.e. independence of physical
quantities on $R$, can be checked {\it a posteriori}, as was done
here. Our calculations showed that the charge and magnetic form
factors of the deuteron are largely insensitive to the value of $R$
for $Q \leq 700$ MeV.  The quadrupole form factor is more sensitive to
the short-distance physics, but is still essentially correct for $Q
\leq 300$ MeV. Indeed, the shape of $F_Q$ is independent of $R$ for $Q
\leq 700$ MeV, but the deuteron's quadrupole moment is somewhat
sensitive to the short-distance potential. The inclusion of a two-body
counterterm for $Q_d$ in the expansion of the electron-deuteron
scattering kernel will remove this undue sensitivity to short-distance
physics.

A consistent, systematic expansion for the electromagnetic current
inside the deuteron is also a result of $\Lambda$-counting. Again,
more complicated effects, including two-body currents, the effects of
nucleon size, and relativistic corrections are suppressed by powers of
$P/\Lambda$.  Indeed, we showed that the impulse approximation with
nucleons of charge $e$ and zero, and magnetic moments $\mu_p$ and
$\mu_n$ is the next-to-leading-order deuteron current in
$\Lambda$-counting.  Corrections are suppressed by powers of
$P^2/\Lambda^2$. We also found that the dominant corrections to this
picture arise from the effects of the finite size of the nucleon, and
from two-body currents. The leading two-body currents in this
expansion have already been found to have a significant effect in
potential-model calculations of electron-deuteron scattering. Thus it
seems that $\Lambda$-counting can be used to systematize some of the
findings of these calculations.

The approach adopted in this work  allows for an accurate
calculation of the long-distance quantities $\langle r^{2n} \rangle$.
These are dominated by the tail of the wave function, which is
precisely reproduced in our approach. Indeed, the deuteron is a
shallow bound state, and so {\it most} of the wave function is in the
tail.  This observation suffices to show that the leading relative
error in $\langle r^{2n} \rangle$ varies as $(\gamma R)^{2n+1}$.  By
contrast, for the electromagnetic deuteron form factors $F_C$, $F_M$
and $F_Q$. the error estimate is only that our computation is accurate
to $O(P^2/\Lambda^2)$. This leads us to observe that among the
corrections arising at orders beyond that considered here, some are
more important than others. For instance, relativistic corrections are
suppressed by powers of $\gamma/M$ and $2/(MR)$, and so turn out to be
less important than corrections due to finite-nucleon size---at least
once $Q^2$ is moderately large.  This is a generic issue in effective
field theory: the error estimate is always conservative. If the
relative error is $(P/\Lambda)^2$ then the estimate is made by taking
the largest possible scale for $P$ and the smallest possible scale for
$\Lambda$. One example of this is electron-deuteron scattering at low
$Q$. Blind application of $\Lambda$-counting suggests an error of size
$(m_\pi/\Lambda_\chi)^2$, but in fact the error in $F_C$ for $Q \lsim
2 \gamma$ is much smaller than this, varying like $Q^2 \gamma R^3$.
Thus a sacrifice is made: efforts to make the power-counting simple
may lead to a significant overestimate of the error of the
calculation.

Of course, $\Lambda$-counting is not the only approach which has been
proposed to EFT in nuclear physics. In Ref.~\cite{Ka98C} the
electromagnetic deuteron form factors were calculated using the
Q-counting scheme developed in Refs.~\cite{Ka98B,Ka98A}.  As far as
direct error estimation goes Q-counting enjoys a significant
practical advantage over $\Lambda$-counting.  In Q-counting all
physical amplitudes are expanded in $Q/\Lambda$ with
\begin{equation}
  Q \sim \gamma, \, m_\pi, \, p;
\end{equation} 
where $p$ is either a nucleon momentum or the momentum of an external
probe.  Since Q-counting is an expansion for amplitudes themselves one
has immediate formal error estimates.  A calculation of a quantity to
order $Q^n$ beyond leading order implies an error of relative order
$(Q/\Lambda)^{n+1}$.  The EFT error estimate then is given by choosing
the largest of $\gamma$, $p$, and $m_\pi$ as $Q$ and the smallest
possible short-distance scale for $\Lambda$.

Within their common domain of applicability, employing
$\Lambda$-counting to derive an $NN$ potential $V$ which is valid to
order $(P/\Lambda)^n$ and then using the Lippmann-Schwinger equation
for the $NN$ amplitude to get $T$ from $V$ is no worse than expanding
$T$ to order $(Q/\Lambda)^n$. Looked at this way it becomes clear that
$\Lambda$-counting includes additional contributions beyond the order
to which one is working in Q-counting. Empirically, we see in both
this and other work that calculations based on $\Lambda$-counting
describe the data no worse than Q-counting calculations with the same
number of parameters. Indeed, they often do better. For example. the
deuteron static properties calculated here and in Ref.~\cite{Pa98} are
considerably more accurate than those of Ref.~\cite{Ka98C}, yet all
three calculations are to next-to-leading order. Similarly, here the
deuteron form factors are well explained to considerably higher $Q$
than in Ref.~\cite{Ka98C}. This suggests that the deuteron wave
function is more accurately described at $r \sim 1/m_\pi$ if
$\Lambda$-counting is employed. The physics which drives this
improvement in the description of the form factors at $Q \sim 2 m_\pi$
is apparently $\Lambda$-counting's use of a OPEP which is iterated to all
orders.

\begin{figure}[h,t,b,p]
  \vspace{0.5cm} \epsfysize=11 cm \centerline{\epsffile{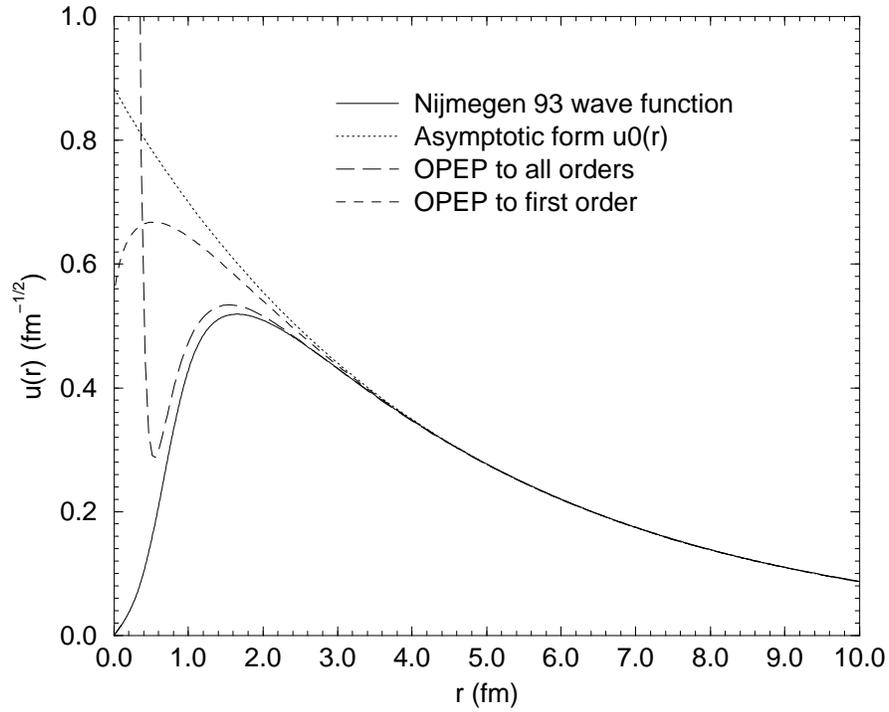}}
  \centerline{\parbox{11cm}{\caption{\label{fig-pertOPE} The S-wave
        radial wave function of the deuteron, $u(r)$. The Nijm93 wave
        function is the solid line, the asymptotic form integrated in
        to $r=0$ without one-pion exchange gives the dotted line,
        while including one-pion exchange when integrating in yields
        the long-dashed line.  The short-dashed line is the result
        when the central part of one-pion exchange is included only to
        first order in perturbation theory.}}}
\end{figure}

In an attempt to elucidate this issue we have integrated the S-wave
deuteron wave function in from $r=\infty$, including the effects of
the central part of one-pion exchange only to leading order in
perturbation theory. The results thereby obtained are quite different
from the nonperturbative~treatment of one-pion exchange employed in
Section~\ref{sec-wavefunction} (see Fig.~\ref{fig-pertOPE}).  Indeed
the resulting $u(r)$ differs significantly from the ``true'' wave
function at $r \sim 2.5$ fm---well below the scale where the wave
function with OPEP iterated to all orders starts to deviate markedly
from $u(r)$ for the Nijm93 potential. The dominant effect which brings
the short-dashed curve of Fig.~\ref{fig-pertOPE} down to the
long-dashed one is the tensor coupling of the D-wave piece of the
deuteron back into the S-wave.  This may well be an indication that
iterating parts of the low-order EFT $NN$ potential (such as one-pion
exchange) to all orders generates effects which are numerically
important even though they are formally of higher-order in Q-counting.

This possibility is not unreasonable since the dimensionless parameter
identified with iterations of the one-pion exchange~ interaction in
S-wave scattering is~\cite{Ka98B}:
\begin{equation}
  \frac{m_\pi}{\Lambda_{NN}} \, = \, \frac{g_A^2 M m_\pi}{16 \pi
    f_\pi^2} \approx 0.5,
\end{equation}
a fairly large expansion parameter~\cite{Ge98C}. This is of particular
concern in the triplet channel, since there are large Clebsch-Gordon
coefficients in the S-D coupling OPEP. In addition, the form of the
tensor force gives an enhancement factor relative to the central force
of 7 at $r=1/m_\pi$.  These large enhancements underlie the
conventional nuclear folklore that the second-order tensor force is
extremely important, indeed far more important than central OPE. In
contrast, in Q-counting, the effects of the tensor force on, say the
${}^3S_1$ phase shift only appear at next-to-next-to-leading order.
This raises questions about how efficiently Q-counting accounts for
these effects.  In contrast, $\Lambda$-counting includes this physics
in a low-order calculation. While none of this demonstrates that
Q-counting fails it sheds some doubt on the desirability of treating
one-pion exchange~ effects perturbatively. A calculation of $NN$ phase
shifts in the ${}^3S_1-{}^3D_1$ channel to NNLO in Q-counting is
near completion~\cite{MS99}.  Clearly, such further
investigation of the efficacy of a perturbative expansion for pionic
physics is very important.

What is also clear is that $\Lambda$-counting leads to an effective
field theory approach for electromagnetic deuteron properties which is
viable for momenta $Q$ up to at least 700 MeV. The picture that
emerges is not very different from that seen in $NN$ potential model
calculations of the same quantities. The main advantage of an
effective field theory treatment at these momentum transfers is 
that corrections to both the $NN$ interaction and the deuteron current
appear as systematic corrections in an expansion in
\begin{equation}
\frac{(p,m_\pi,Q)}{(\Lambda_\chi,M)},
\end{equation}
where $p$ is the typical momentum inside the deuteron and $\Lambda_\chi$
is the scale of chiral symmetry breaking. 

\section*{Acknowledgments}

D.~R.~P. gratefully acknowledges useful discussions with Mart
Rentmeester and Vincent Stoks. We both thank Silas Beane, Martin
Savage, and Bira van Kolck for stimulating conversations on this
subject, and for helpful comments on the manuscript. Thanks are also
due to Vincent Stoks for supplying us with the Nijm93 wave function.
We are both grateful for the support of the U.~S.  Department of
Energy, Nuclear Physics Division (grants DE-FG02-93ER-40762 and
DE-FG03-970ER-41014).

\end{document}